\documentclass[%
 reprint,
superscriptaddress,
showpacs,preprintnumbers,showkeys,
footinbib,
bibnotes,
 amsmath,amssymb,
 aps,
]{revtex4-1}

\usepackage{graphicx}% Include figure files
\usepackage{dcolumn}% Align table columns on decimal point
\usepackage{bm}% bold math
\usepackage{xcolor}% color text/math
\usepackage{ bbold }% \mathbb{}
\usepackage{acronym}

\usepackage{subfigure,amsmath}
\newacro{goe}[GOE]{Gaussian orthogonal ensemble}
\newacro{FOE}[FOE]{Fluctuation operator expansion}
\newacro{P}[P]{Poisson}
\newacro{ME}[ME]{mobility edge}
\newacro{MF}[MF]{mean-field}
\newacro{QP}[QP]{quasiparticle}
\newacro{MBL}[MBL]{many-body localization}
\newacro{BG}[BG]{Bose glass}
\newacro{BHM}[BHM]{Bose-Hubbard model}
\newacro{BKT}[BKT]{Beresinskii-Kosterlitz-Thouless}
\newacro{LIOM}[LIOM]{local integral of motion}
\newacro{SF}[SF]{superfluid}
\begin{document}

%\preprint{APS/123-QED}

\title{Mobility edge of the two dimensional Bose-Hubbard model}% Force line breaks with \\

\author{Andreas Gei{\ss}ler}
\email{andreas.geissler87@gmail.com}
\affiliation{icFRC, ISIS, University of Strasbourg and CNRS, 67000 Strasbourg, France}
\affiliation{Institut f\"ur Theoretische Physik, Goethe-Universit\"at, 60438 Frankfurt/Main, Germany}
%\author{Johannes Schachenmayer}
%\affiliation{ISIS, University of Strasbourg and CNRS, 67000 Strasbourg, France}%
\author{Guido Pupillo}
\affiliation{icFRC, ISIS, University of Strasbourg and CNRS, 67000 Strasbourg, France}

\date{\today}% It is always \today, today,

\begin{abstract}

We analyze the disorder driven localization of the two dimensional Bose-Hubbard model by evaluating the full low energy quasiparticle spectrum via a recently developed fluctuation operator expansion method. For any strength of the local interaction we find a mobility edge that displays an approximately exponential decay with increasing disorder strength. We determine the finite-size scaling collapse and exponents at this critical line finding that the localization of excitations is characterized by weak multi-fractality and a thermal-like critical gap ratio. A direct comparison to a recent experiment yields an excellent match of the predicted finite-size transition point and scaling of single particle correlations.
 
\end{abstract}

% PACS
%Use showkeys class option if keyword display desired
                              
\maketitle

%\tableofcontents

\section{Introduction}

In the last decade the study of disorder-driven localization of quantum particles has received considerable interest, following the suggestion that  Anderson localization for non-interacting models \cite{Anderson1958,Abrahams1979,Fleishman1980,Roati2008} can be generalized to interacting ones \cite{Altshuler1997,Basko2006,Oganesyan2007,Pal2010} in the framework of the so-called many-body localization (MBL). One of the most prominent features of MBL  is its incompatibility with the eigenstate thermalization hypothesis (ETH) resulting from an extensive number of local integrals of motion \cite{Serbyn2013a,Huse2014,Chandran2015a,Ros2015,Imbrie2017}. A complete demonstration of MBL would in principle require knowledge of the whole spectrum, limiting the use of exact diagonalization techniques to small system sizes, especially when bosonic particles are considered \cite{Sierant2018,Wahl2019}. The existence of MBL has been rigorously proven in one-dimensional (1D) spin-chains \cite{Imbrie2016,Imbrie2016a}, while various perturbative arguments \cite{Fleishman1980,Altshuler1997,Basko2006,Nandkishore2014} and numerical evidence \cite{Oganesyan2007,Pal2010,Kshetrimayum2019} have also supported its existence in two dimensions - involving a mobility edge (ME) separating mobile from localized states in the spectrum. However, recent theoretical arguments have challenged the existence of MBL both in 1D \cite{Suntajs2019}  and 2D \cite{DeRoeck2016,Agarwal2017,DeRoeck2017} in the thermodynamic limit. Experimental realizations of bosonic systems have already been achieved in cold atom setups where a disorder potential can be imprinted onto a confined optical lattice in 1D \cite{Lukin2019,Rispoli2018} and 2D \cite{Choi2016,Rubio-Abadal2019}, showing strong signs of high energy localization in confined systems for both cases. Related experiments \cite{Fallani2007,Meldgin2016} have also observed evidence for a ground state Bose-glass phase compatible with theoretical predictions of a zero-energy superfluid to Bose-glass transition \cite{Fisher1989,Makivic1993,Zhang1995,Priyadarshee2006,Soyler2011, AlvarezZuniga2013,Saliba2014,AlvarezZuniga2015}.\begin{figure}[t]
 \centering
 \includegraphics[width=0.99\columnwidth]{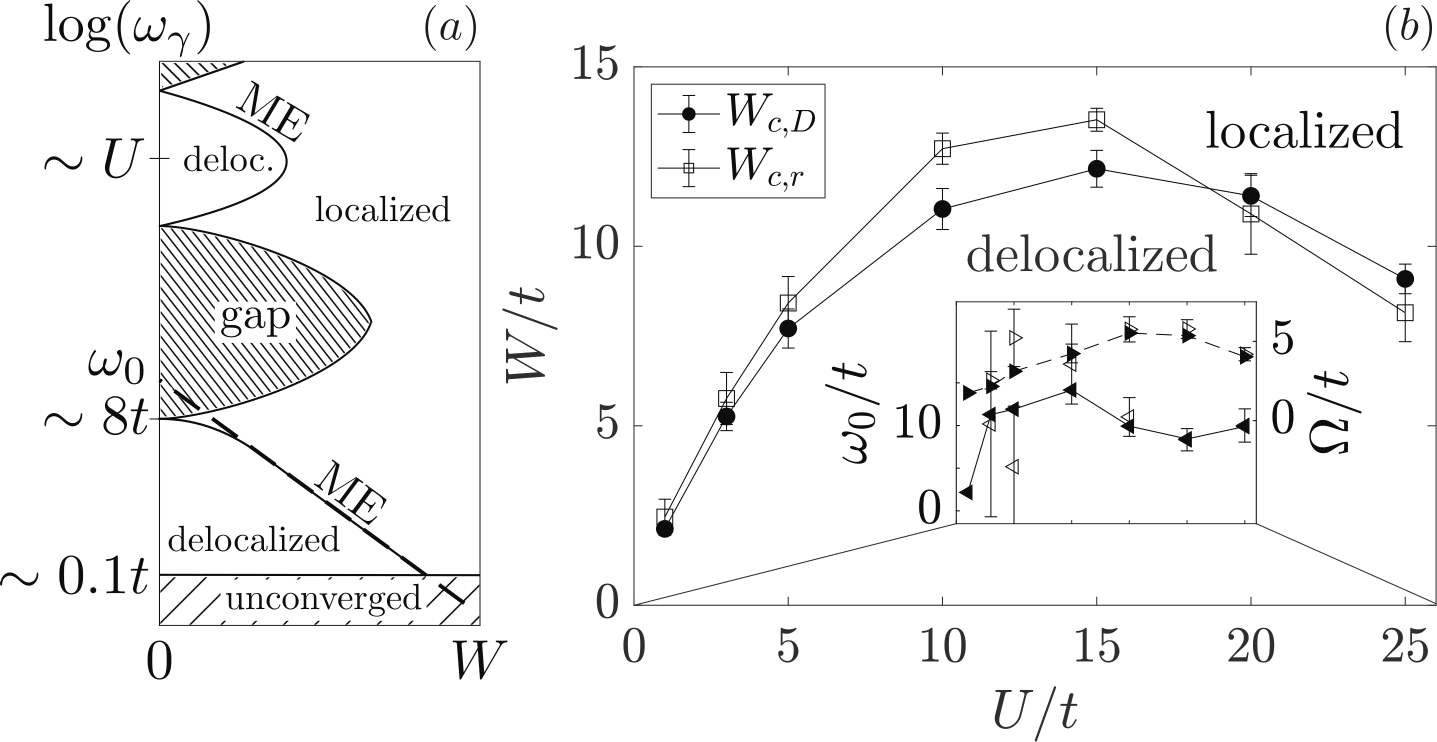}
 \caption{Quasiparticle localization in the disordered 2D Bose-Hubbard model~\eqref{eq:MBL-hamil}.
$(a)$: Typical structure of the quasiparticle spectrum at fixed interaction $U>8t$. For sufficiently weak disorder $W$ a band gap persists, while multiple MEs centered at the homogeneous bands separate localized from delocalized states. The lowest ME displays exponential behavior (dashed line) down to the lowest resolved QP energies.
$(b)$: ME $W_c(\omega)$ of excitations at a quasiparticle energy of $\omega = t$ as a function of $U/t$, determined  from the fractal dimension $D$ and the gap ratio $r$ (see legend). \textit{Inset:} Amplitude $\omega_0$ (left pointing triangles) and decay constant $\Omega$ (right pointing triangles) characterizing the lowest ME~\eqref{eq:ME_exp}.
 }
 \label{fig:MBL_phases}
\end{figure}

Here, we investigate localization effects in the excitation spectrum of the two-dimensional Bose-Hubbard model (BHM) in the presence of disorder utilizing a recently developed fluctuation operator expansion (FOE) method  \cite{Bissbort2014,Geissler2018}, which gives access to the complete spectrum of quasiparticle (QP) excitations for system sizes comparable to experiments. Our results are summarized in Fig.~\ref{fig:MBL_phases}. For all interaction strengths disorder induces (at least) one ME. We determine the finite-size scaling at the critical points characterized by weak fractality and a thermal-like critical gap ratio. Importantly, in the limits of our numerical QP method the low energy ME converges onto an exponential decay with disorder.
For the case of particles confined by a harmonic potential, we compute correlation functions and extract the inverse decay length, finding excellent agreement with recent experiments \cite{Choi2016} in terms of a finite-size localization transition.

In the following we first introduce the model (Set.~\ref{sec:model}) and the fluctuation operator expansion (Sec.~\ref{sec:FOE}). In Sec.~\ref{sec:characterization} we introduce the observables used to characterize localization and discuss their finite-size collapse determining the ME (Sec.~\ref{sec:scaling}). We determine the finite-size localization transition of a harmonically trapped system (Sec.~\ref{sec:trapMBL}) and directly compare to the experimental results of reference \cite{Choi2016}. In Sec.~\ref{sec:conclusion} finally, we end with some concluding remarks.

\section{Model}\label{sec:model}

The Hamiltonian of the BHM with on-site disorder and in the grand canonical ensemble reads
\begin{align} \label{eq:MBL-hamil}
\hat{H} =  \sum_{\ell}^{L^2} \underbrace{ \left( \mu_{\ell} \hat{b}^{\dag}_{\ell} \hat{b}_{\ell} + \frac{U}{2} \hat{b}^{\dag}_{\ell} \hat{b}^{\dag}_{\ell} \hat{b}_{\ell} \hat{b}_{\ell} \right) }_{\hat{H}_{\ell}} - t \sum_{\langle \ell,\ell' \rangle} (\hat{b}^{\dag}_{\ell} \hat{b}_{\ell'} + \textrm{h.c.} ),
\end{align}
where $\hat{b}^{\dag}_{\ell}$ $(\hat{b}_{\ell})$ are bosonic creation (annihilation) operators at the site $\ell$, $t$ is the
tunneling rate between nearest neighbor sites $\langle \ell, \ell '\rangle$ on a square lattice of spacing $a$ and linear size $L$, while $U$ is the local on-site Hubbard interaction. The energy $\mu_{\ell}$ reads $\mu_{\ell} = -\mu + \epsilon_{\ell}$, with $\mu$ the chemical potential fixing the particle number and $\epsilon_{\ell}$ a local energy shift due to disorder or an external harmonic potential.  
With Ref.~\cite{Choi2016} in mind we choose a Gaussian probability distribution $P(\epsilon_{\ell}) = \left( 2 \pi W^2 \right)^{-1/2} \exp\left[ - \epsilon_{\ell}^2 / (2 W^2)  \right]$ with the standard deviation $W$ \footnote{A crucial difference compared to the commonly considered box disorder is the presence of rare extreme peaks or wells in the potential.}. In this work we analyze this model over a range of interactions $U/t \in [1, 25]$ and disorder strengths $W/t \in [1, 15]$ at half-filling. We furthermore investigate the effect of an external trapping potential in order to compare with the recent experiment \cite{Choi2016} for $U = 24.4t$ and $W/t \in [0.4 , 7]$.

\section{Fluctuation operator expansion}\label{sec:FOE}
 
The FOE~\cite{Bissbort2014,Geissler2018} is a QP method based on a Gutzwiller expansion of~\eqref{eq:MBL-hamil} in terms of eigenstates $|i \rangle_{\ell}$ of the local mean-field Hamiltonians $\hat{H}_{\textrm{MF}}^{(\ell)} = \hat{H}_{\ell} - t \sum_{ \lbrace \ell' | \langle \ell, \ell' \rangle \rbrace} \left( \hat{b}^{\dag}_{\ell} \phi_{\ell'} + \textrm{h.c.} \right)$, where the fluctuation operators $\hat{\delta b}_{\ell} \equiv \hat{b}_{\ell} - \phi_{\ell}$ and the fields $\phi_{\ell} \stackrel{!}{=} {}_{\ell}\langle 0| \hat{b}_{\ell} |0 \rangle_{\ell}$ are determined self-consistently.
For $N \rightarrow \infty$, $\hat{\delta b}_{\ell} = \sum_{i,j=0}^N {}_{\ell}\langle i | \hat{\delta b}_{\ell} | j \rangle_{\ell} |i \rangle_{\ell}  {}_{\ell}\langle j|$ constitutes an exact quadratic map onto a complete basis set of the local Gutzwiller raising (lowering) operators $\sigma_{\ell}^{(i)^{\dagger}} \equiv |i \rangle_{\ell}  {}_{\ell}\langle 0|$ ($\sigma_{\ell}^{(i)} \equiv |0 \rangle_{\ell}  {}_{\ell}\langle i|$). These generate arbitrary local fluctuations $\kappa_{\ell} = \sum_{i>0} \sigma_{\ell}^{(i)^{\dag}} \sigma_{\ell}^{(i)}$ of any self-consistent MF state $|\psi_{\textrm{MF}} \rangle = \prod_{\ell} | 0 \rangle_{\ell}$. The quality of the approximation is ascertained for $\kappa = L^{-2} \sum_{\ell} \langle \kappa_{\ell} \rangle \ll 1$  (in this work we always find this criterion to be fulfilled in the quasiparticle ground state~\eqref{eq:qp-groundstate} \footnote{We refer the interested reader to our follow-up work \cite{Geissler2020} for a detailed discussion of the FOE method in the context of disordered systems.}).
Here, we consider terms of second order in the Gutzwiller operators, which using $\pmb{\sigma} = \left( \sigma_{1}^{(1)} , \ldots  , \sigma_{L^2}^{(N)} \right)^{\textrm{T}}$ yields the following approximate representation of $\hat{H}$ \footnote{We note that the MF self-consistency condition guarantees the absence of first order terms.}
\begin{align}
\hat{\mathcal{H}}^{(2)}_{\textrm{QP}} \equiv \frac{1}{2} \begin{pmatrix}
{\pmb{\sigma}} \\ {\pmb{\sigma}}^{\dag}
\end{pmatrix} ^{\dag} \begin{pmatrix}
{h} & {\Delta} \\
{\Delta}^* & {h}^*  
\end{pmatrix} \begin{pmatrix}
{\pmb{\sigma}} \\ {\pmb{\sigma}}^{\dag}
\end{pmatrix} -\frac{1}{2} \textrm{Tr}(h). \label{eq:H2_QP} 
\end{align}
The scalar term $\textrm{Tr}(h)/2$ results from reordering normal ordered terms to anti-normal order,
while the matrix entries are given by the local matrix elements ${B}^{(\ell)}_{i,j} \equiv {}_{\ell}{\langle} i | \hat{b}_{\ell} | j \rangle_{\ell}$,

\begin{align}
{h}_{(i,\ell),(j,\ell')} =& - t_{\ell,\ell'} \left( {B}^{(\ell)*}_{0,i} {B}^{(\ell')}_{0,j} + {B}^{(\ell)}_{i,0} {B}^{(\ell')*}_{j,0} \right) \label{eq:QPmatrix_h}  \\ &+ \delta_{\ell,\ell'} \delta_{i,j} (E_i^{(\ell)}-E_0^{(\ell)}), \nonumber \\
{\Delta}_{(i,\ell),(j,\ell')} =& - t_{\ell,\ell'} \left( {B}^{(\ell)*}_{0,i} {B}^{(\ell')}_{j,0} + {B}^{(\ell)}_{i,0} {B}^{(\ell')*}_{0,j} \right). \label{eq:QPmatrix_Delta}
\end{align}
Here, $t_{\ell,\ell'}$ is the tunneling matrix with nonzero entries only for neighboring sites, and $E_i^{(\ell)}$ are the local excitation energies of the $i$th Gutzwiller excitation at site $\ell$.

The diagonalization of~\eqref{eq:H2_QP} yields 
$\hat{H} \approx {\sum_{\gamma}} \omega_{\gamma} \beta_{\gamma}^{\dag} \beta_{\gamma} + \Delta E_{\textrm{QP}}$ %\label{eq:HQP_diag_full}
in terms of infinitely lived QP modes $\gamma$ with corresponding energies $\omega_{\gamma}$ and $\beta_{\gamma} \equiv \mathbf{u}^{(\gamma)^{\dag}}\boldsymbol{\sigma} + \mathbf{v}^{(\gamma)^{\dag}} \boldsymbol{\sigma}^{\dag}$ are the generalized Bogoliubov-type operators, with $\mathbf{u}^{(\gamma)}$ and $\mathbf{v}^{(\gamma)}$ the corresponding eigenvectors. Analogous to standard Bogoliubov theory, these inherit approximately bosonic commutation relations $[ \beta_{\gamma}, \beta_{\gamma'}^{\dagger} ] \approx \delta_{\gamma,\gamma'}$ from the Gutzwiller operators for $|\mathbf{u}^{(\gamma)}|^2 - |\mathbf{v}^{(\gamma)}|^2 = 1$. $\mathbf{v}^{(\gamma)}$ and $\mathbf{u}^{(\gamma)}$ can be interpreted as dual wave-functions analogous to particle and hole fluctuations. Normal ordering of  operators results in a scalar correction $\Delta E_{\textrm{QP}}$, irrelevant to the present discussion \cite{Geissler2018}. Finally, we implicitly define the QP ground state via
\begin{align} \label{eq:qp-groundstate}
\beta_{\gamma} | \psi_{\textrm{QP}} \rangle = 0 \;  \forall \gamma
\end{align}
which also best fulfills the approximation of neglected QP interactions \cite{Bissbort2014,Geissler2018}.

Drawing from variational concepts \cite{Huber2007,Huber2008,Bissbort2011,Endres2012} and based on a MF description that becomes exact for weak and strong interactions, the FOE allows for a systematic, non-perturbative improvement over standard Bogoliubov theory \cite{Bogolyubov1947} that also incorporates effects of many-body entanglement \footnote{See Supplemental Material at [] for a discussion of details of the finite-size scaling analysis for the excitations and the ground state, a comparison of scenarios for the decay of correlations in a trapped system and a discussion of the entanglement captured by the FOE method.}. It gives access to the otherwise neglected gapped (amplitude) Hubbard subbands in the disorder-free limit of $ \hat{H}$ in Eq.~\eqref{eq:MBL-hamil}~\cite{Bissbort2014,Geissler2018}.
As we show in the following sections, these modes, absent in standard Bogolioubov theory, play an important role in the localization transition at finite energy. We note that Bogoliubov quasiparticle theory has already been used  to successfully investigate  2D localization at low energy (e.g. \cite{AlvarezZuniga2013,Saliba2014,Gaul2015}), and in particular the existence of a Bose-Glass phase for hard-core bosons (i.e., $U\rightarrow \infty$) with binary disorder \cite{AlvarezZuniga2013}. Numerous works have unambiguously demonstrated the existence of a direct zero-energy phase transition between a Bose condensed superfluid and a Bose-Glass for the 2D BHM with uniform disorder distribution, similar to Eq.~\eqref{eq:MBL-hamil}~\cite{Fisher1989,Makivic1993,Zhang1995,Priyadarshee2006,AlvarezZuniga2015}. Here we focus on the existence of a finite-energy ME.

\section{Localization characteristics}\label{sec:characterization}

\begin{figure}[t]
 \centering
 \includegraphics[width=0.99\columnwidth]{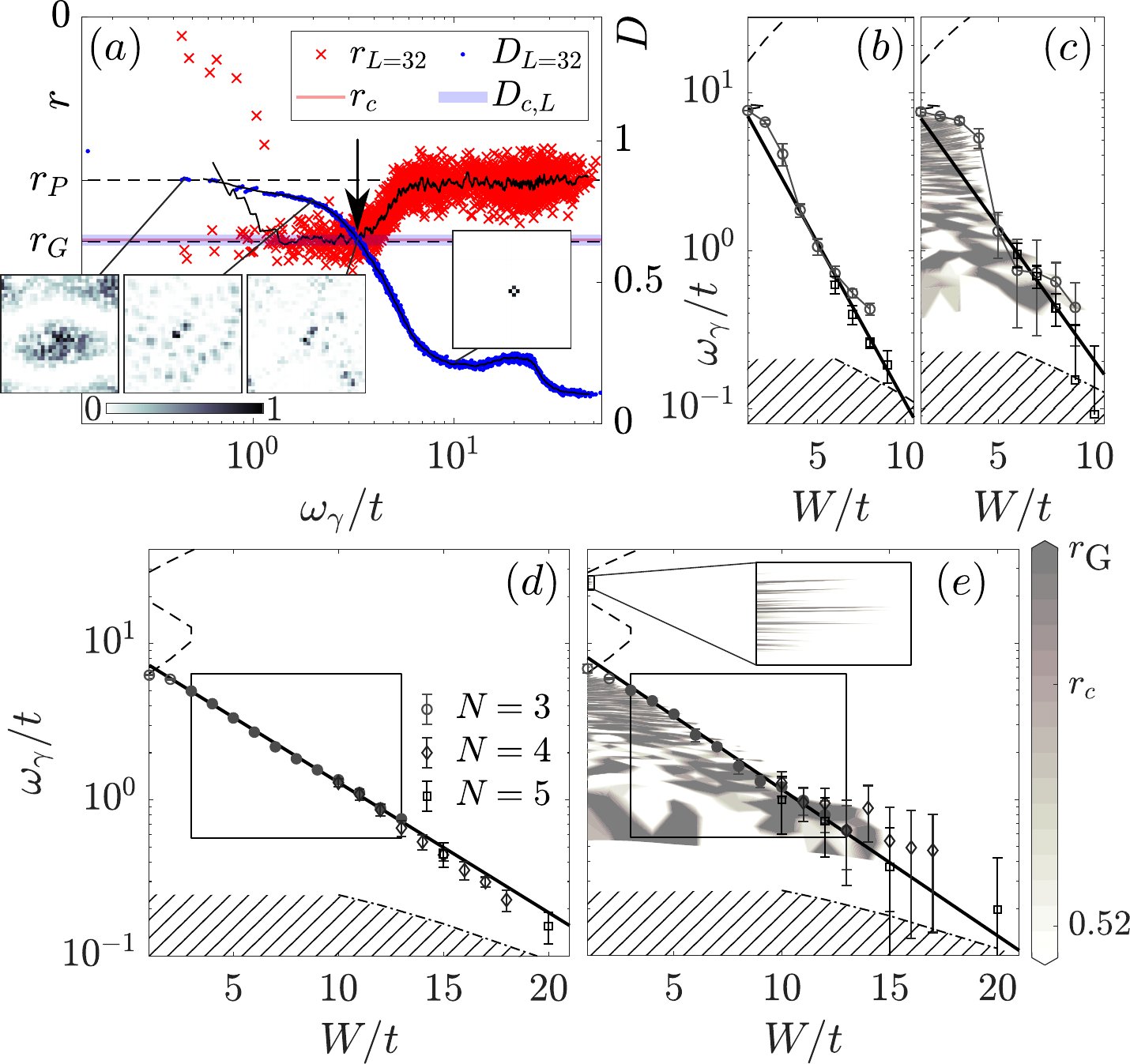}
 \caption{$(a)$: Gap ratio $r$ (left ordinate, inverted) and fractal dimension $D$ data (right ordinate) as functions of the QP energy $\omega_{\gamma}/t$ for $U/t = 20$, $W/t=5$ and $L=32$ averaged over 95 realizations. Black lines are moving averages (of 21 points) as a guide to the eye and dashed lines mark $r_P$ and $r_{G}$. The crossing point (vertical arrow) of the data with the critical $r_c$ (shaded red, narrow) and $D_{c,L}$ (shaded blue, wide) mark the ME. \textit{Insets}: Exemplary squared QP wave-functions $|\mathbf{v}_{\ell}^{(\gamma)}|^2$ with maxima normalized to one. $(b$-$e)$: Separation of the QP spectra by the ME for $U/t = 3$ $(b,c)$ and $U/t = 20$ $(d,e)$. Dashed lines mark band edges, dash-dotted lines signify lowest resolved energies for $N=5$ (dashed regions, see Supp. Mat.), while data points mark the ME with the respective FOE truncation given in the legend of $(d)$. Panels $(c,e)$ are contour plots of $r$ [color scale in $(e)$] binned under the condition $r>0.3$ (see text). Inset $(e)$ shows remnants of a ME for the upper bands, while large boxes in $(d,e)$ mark the region for which finite-size scaling has been performed, yielding the filled data points. Thick black lines in $(b-e)$ are fits of eq.~\eqref{eq:ME_exp} (see text).}
 \label{fig:colormap_rANDdf}
\end{figure}

To characterize the degree of localization we consider the following two observables: (i) The gap ratio 
\begin{align}
r_{\gamma}\equiv \left\langle \frac{\textrm{min}[\Delta \omega_{\gamma-1},\Delta \omega_{\gamma}]}{\textrm{max}[\Delta \omega_{\gamma-1},\Delta \omega_{\gamma}]} \right\rangle_d
\end{align}
with  $\Delta \omega_{\gamma} = \omega_{\gamma+1} - \omega_{\gamma}$ the quasiparticle energy gaps and $\langle \cdot \rangle_d$ the disorder average. The observable $r_{\gamma}$ is known from random matrix theory \cite{Oganesyan2007,Atas2013} to have the mean value $ r_{\textrm{G}} \approx 0.5307$ and $ r_{\textrm{P}}= 2 \textrm{ln} 2 - 1 \approx 0.3863$ in the delocalized and localized phases, respectively, resulting from level statistics belonging to the Gaussian orthogonal and Poisson ensembles. The second observable is (ii) the fractal dimension $D^{(\gamma)}_{q=2}$ of the QP fluctuation wave-functions $\mathbf{v}^{(\gamma)}$. Analogous to the scaling of $q$-moments $R_q=\sum_n \left| \psi_n \right|^{2q}$ of many-body eigenstates~ \cite{Hentschel1983,Mace2018,Lindinger2019} we define\begin{align}
 {D_{q=2}^{(\gamma)}} = \left\langle  -\log_{L^2} \frac{\sum_{\ell}^{L^2} |\mathbf{v}^{(\gamma)}_{\ell}|^{4}}{\sum_{\ell}^{L^2} |\mathbf{v}^{(\gamma)}_{\ell}|^2} \right\rangle_d, \label{eq:fractal_dim}
\end{align}
for the local amplitudes $|\mathbf{v}^{(\gamma)}_{\ell}|^2 = \sum_{i>0} |\mathbf{v}^{(\gamma)}_{\ell,i}|^2$ of the wave-function, which naturally characterize the spatial extension of each QP mode [see examples in Fig.~\ref{fig:colormap_rANDdf}(a)]. For our purposes we consider $q=2$, while one obtains the multifractality spectrum by also taking all other values $q>0$ into account.

Delocalized states with $r \approx  r_{\textrm{G}}$ \footnote{From here on we omit the indices of the observables unless they are necessary.} appear primarily at low QP energies $\omega_{\gamma} / t$ and for weak disorder $W/t$, as shown in the contour plots Fig.~\ref{fig:colormap_rANDdf}$(c,e)$ for weak ($U = 3t$) and strong ($U = 20t$) interactions, respectively. We note that values of $r < r_P$ for weak disorder and small QP energies [e.g. for $\omega_{\gamma} \lesssim t$ in Fig.~\ref{fig:colormap_rANDdf}($a$)] result from symmetry related finite-size effects irrelevant to our discussion. 

For $U/t \gtrsim 20$ and $W/t\lesssim 1$ we find a band of additional delocalized states for energies $\omega_\gamma \sim U$ reflecting the presence of typical Hubbard subbands which overlap for $U \lesssim t$ [dashed lines in Fig.~\ref{fig:colormap_rANDdf}$(b-e)$]. In all cases, increasing $W/t$ spreads the bands so they overlap and drives a transition to localized states with $r_{\textrm{G}} \gnsim r \geq r_{\textrm{P}}$, implying the existence of (multiple) MEs. For the same cases we find similar behavior for the fractal dimension $D$ down to the truncation limit [compare Fig.~\ref{fig:colormap_rANDdf}$(b,d)$]. Data points in Figs.~\ref{fig:colormap_rANDdf}$(b-d)$ mark the MEs determined via a finite-size scaling as discussed in the next section.

\subsection{Finite-size scaling analysis}\label{sec:scaling}

\begin{figure}[t]
 \centering

\includegraphics[width=0.99\columnwidth]{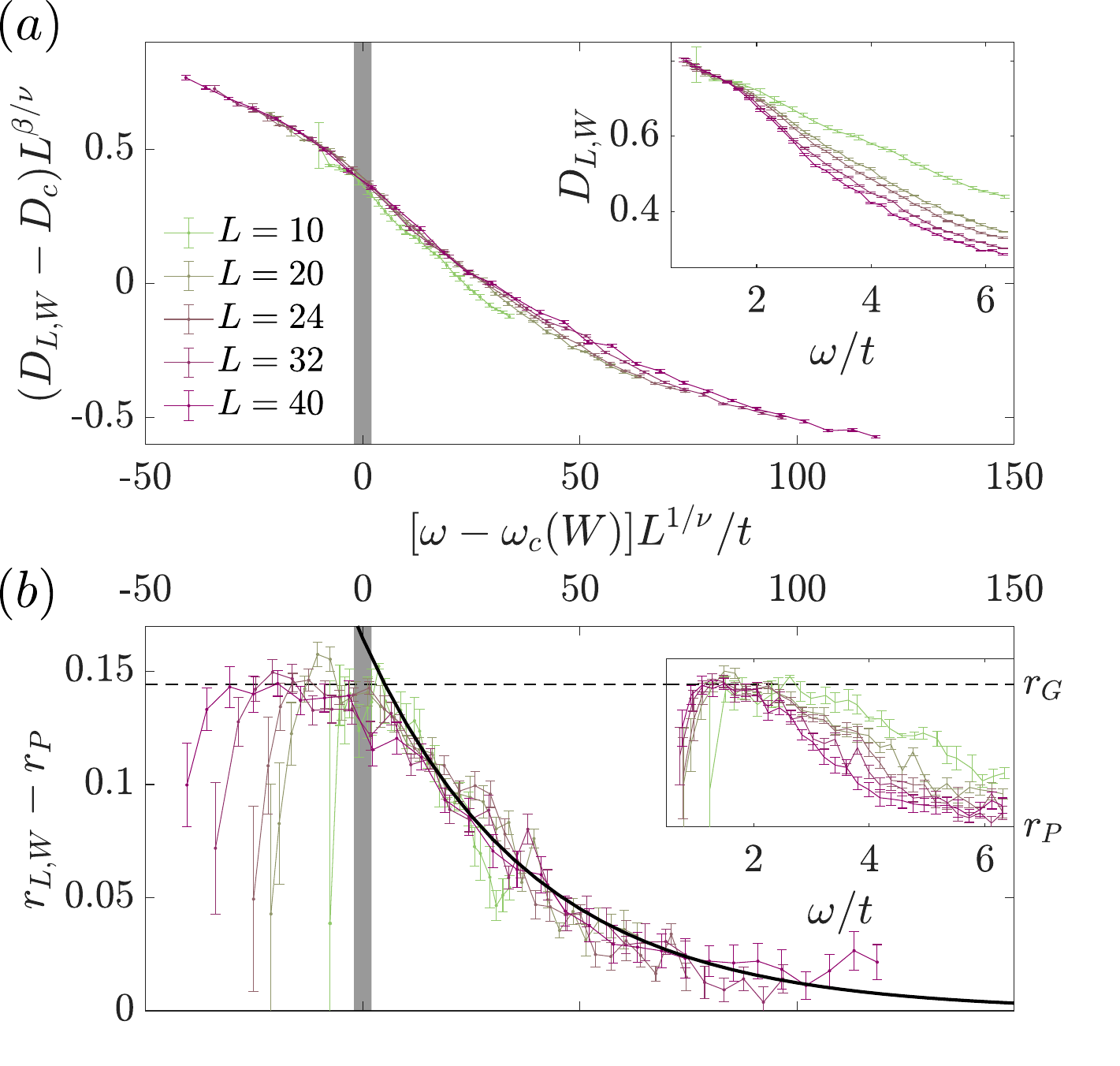}
 \caption{
Exemplary scaling collapse of $D$ $(a)$ and $r$ $(b)$. For the finite-size scaling every data set is binned for 30 equal spaced energies [within the region shown in Fig.~\ref{fig:colormap_rANDdf}$(d,e)$], while $L \in [10,20,24,32,40]$ [legend in $(a)$] with bins containing $[4, 8, 12, 20, 32]$ disorder averaged data values closest in $\omega_{\gamma}$ to $\omega$, respectively, and $W/t = 7$. In $(b)$ the horizontal dashed line marks $r_{G}-r_P$, the solid line is an exponential fit as guide for the eye and data within the grey shaded regions is used to determine $r_c$ and $\langle \tilde{D}_W(0) \rangle_{ME}$. Insets show unscaled data.
}
 \label{fig:alg_scaling}
\end{figure}

We determine the position of the (lowest energy) ME via finite-size scaling for the case $U = 20t$, with linear sizes $L \in \lbrace 10,20,24,32,40 \rbrace$ and corresponding numbers of realizations $N_r \in \lbrace 480,240,240,95,48 \rbrace$ for $N=3$ which we find to be sufficient here \cite{Note4}. We find the data to be consistent with the scaling relations
\begin{align}
r_{L,W}(\omega) &= \tilde{r}_W \left(\left[\omega-\omega_c(W)\right]  L^{1/\nu} \right) , \\
D_{L,W}(\omega) - D_c &= L^{-\beta/\nu} \tilde{D}_W\left(\left[\omega-\omega_c(W)\right] L^{1/\nu}\right).
\end{align}
Here, $\tilde{r}_W(\cdot)$ and $\tilde{D}_W(\cdot)$ are the scaling functions, while the universal scaling exponents $\lbrace \beta, \nu \rbrace$ and the critical fractal dimension $D_c$ are to be determined self-consistently in combination with the critical energies $\omega_c(W)$ corresponding to the ME.
Figures~\ref{fig:alg_scaling} show exemplary data collapses of $D$ [panel $(a)$] and $r$ [panel $(b)$] over the QP energies, $W/t = 7$ and all system sizes $L$, where collapses have been performed for all the data in the region within the large black boxes in Fig.~\ref{fig:colormap_rANDdf}$(d,e)$ with filled symbols marking the scaling result \cite{Note4}. As a result of all collapses we find 
\begin{align}
\beta/\nu &= 0.26(5), & 1/\nu&=0.91(4), & D_c&=0.51(3),
\end{align}
implying weak fractal behavior at the critical point. While we get a good collapse for each individual disorder value (Fig.~\ref{fig:alg_scaling} and Supp. Mat.) deviations from a single line imply a weak dependency of $\tilde{D}_W$ and $\tilde{r}_W$ on $W$. Also, the decay of $r$ towards $r_{P}$ is always nearly exponential [black line Fig.~\ref{fig:colormap_rANDdf} $(c)$, and Supp. Mat.].
From the collapsed data at the critical point we extract a thermal-like $r_{c}=\langle\tilde{r}_W(0)\rangle_{W}=0.527(3) \approx r_G$ consistent with the weak fractality of the critical QP states and $\langle \tilde{D}_W(0) \rangle_{W} = 0.35(3)$. Here, $\langle \cdot \rangle_{W}$ is the average over $W$ inside the large boxes in Fig.~\ref{fig:colormap_rANDdf}$(d,e)$.

\medskip
Next, we determine two independent estimates of $\omega_c(W)$ for other $U/t$ at fixed $L=32$ and up to $N=5$, which is necessary to determine the low energy ME at strong disorder. We take the crossing points of (i) $D$-data with the finite-size critical dimension $D_{c,L=32}=0.657(16)$ [Fig.~\ref{fig:colormap_rANDdf}$(a)$, black arrow], as well as of (ii) exponential fits to $r$-data with the critical gap ratio $r_c$ [see black line in Fig.~\ref{fig:alg_scaling}$(b)$ and Fig.~\ref{fig:colormap_rANDdf}$(a)$, black arrow] \footnote{Most of the time the exponential gap ratio fits slightly overestimate the critical energy, but always well within the errorbars.}.
For $U/t \in \{3,20 \}$, respectively, Figs.~\ref{fig:colormap_rANDdf} show the $D_{c,L=32}$ MEs [empty symbols, panels $(b,d)$] and binned $r$-data [6 values per bin, panels $(c,e)$] close to the critical $r_c$. We note that $L=40$ for $N=5$ in panels $(d,e)$ is neccessary to resolve the low energy ME as we discuss in the Supp. Mat.
Excitingly, this procedure leads to consistent values for $\omega_c(W)$ for all considered values of $W$ and $U$. 
Interestingly, we find that for all data sets and sufficiently small $\omega$, the dependence of $\omega_c(W)$ on $W$ is consistent with the empirical ansatz
\begin{align}
\omega_c(W) = \omega_0 \exp (-W / \Omega) \label{eq:ME_exp},
\end{align}
except for small $U$ where the gap to the upper band already vanishes at small $W$.
Corresponding exponential fits to the $N=3$ data, shown as thick black lines in Figs.~\ref{fig:colormap_rANDdf}($b$-$e$), work well in a large part of the spectrum, while additional data obtained by increasing $N$ and $L$ matches up perfectly for disorder values beyond the $N=3$ truncation limit \cite{Note4}. Panel $(b)$ of Fig.~\ref{fig:MBL_phases}$(b)$ summarizes these findings, showing the extension of delocalized QP states up to the ME $W_c(\omega)$ as a function of interaction at fixed energy $\omega=t$  with its greatest extension at $U/t \approx 15$, while the parameters of~\eqref{eq:ME_exp} are given in the inset of Fig.~\ref{fig:MBL_phases}$(b)$ depicting amplitudes $\omega_0$ and decay constants $\Omega$ as functions of $U/t$. We note that the perfect match of~\eqref{eq:ME_exp} for increased truncation and system sizes implies the absence of a thermal to fully QP localized phase, if extended to the thermodynamic limit \cite{DeRoeck2016,Agarwal2017,DeRoeck2017}.

\section{Trapped system}\label{sec:trapMBL}

\begin{figure}[t]
 \centering

 \includegraphics[width=0.99\columnwidth]{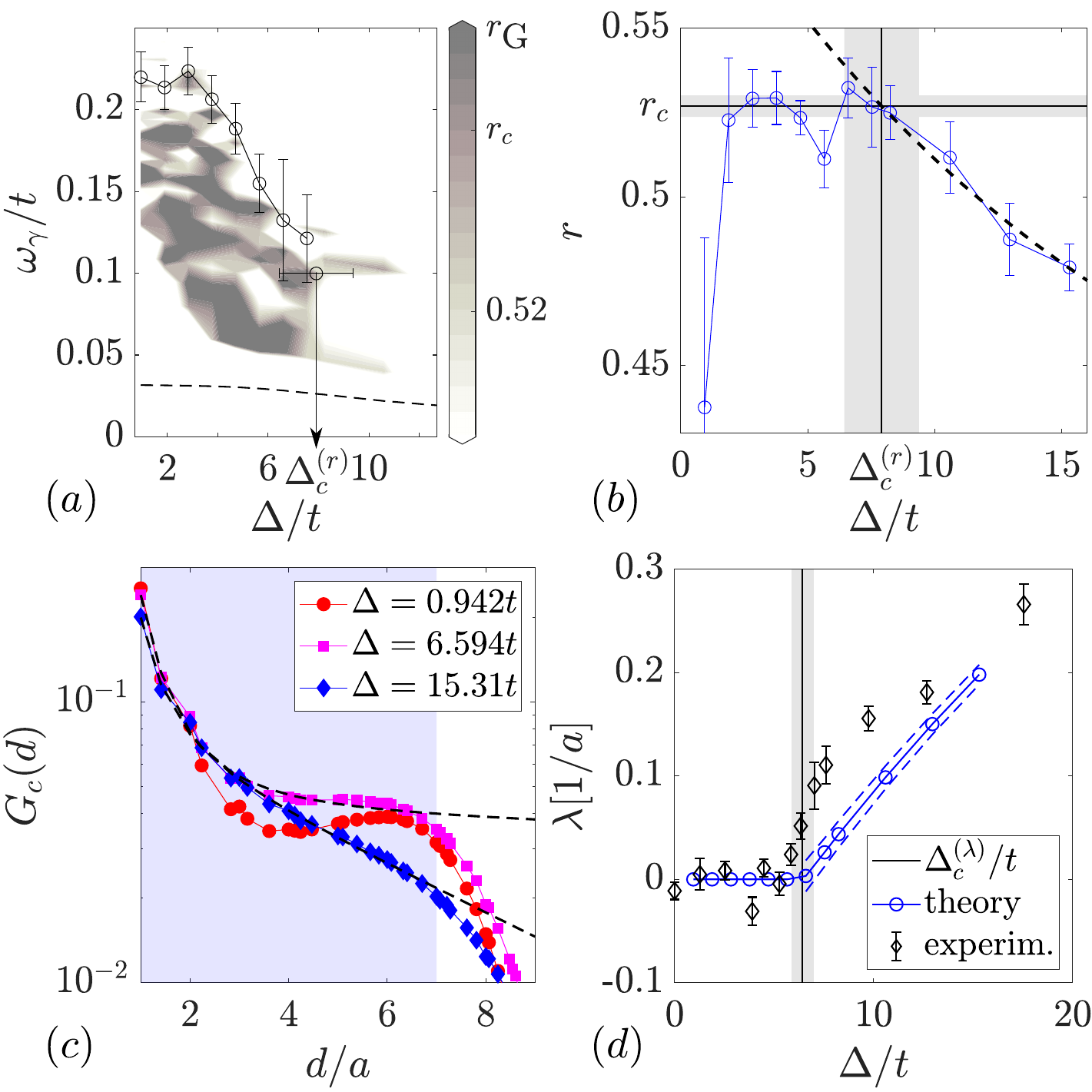}

 \caption{
Localization in a harmonically trapped lattice. $(a)$: Binned (every 6 gap pairs) gap ratio contours of the QP spectra as function of $\Delta/t$. Circles mark the ME obtained from the crossing of an exponential fit to each $r(\omega_{\gamma})$ with $r_c$ and the dashed line marks the lowest resolved QP excitation. $(b)$: Circles are binned data for the 11 $r$ values closest to $\omega_{\gamma}/t = 0.1$ [between solid lines in $(a)$]. The crossing of the exponential fit (dashed line) with $r_c$ yields $\Delta^{(r)}_c$ [vertical arrow in $(a)$], both shown as black lines together with associated errors. $(c)$: Exemplary fits of~\eqref{eq:trap_Greens} to the numerical $G_c$ for various $\Delta$. $(d)$: Inverse decay length $\lambda$ of~\eqref{eq:trap_Greens} in comparison to experimental data \cite{Choi2016}. The black solid line marks the theoretical prediction of $\Delta_c^{(\lambda)}$ together with one standard deviation.
}
 \label{fig:trap_results}
\end{figure}

We end our discussion with the analysis of the added effect of a harmonic trap as realized in \cite{Choi2016} approximating the skewed Gaussian disorder used therein by an exact Gaussian with the full width at half maximum $\Delta = 2 \sqrt{2 \textrm{ln} 2} W$. All other parameters of~\eqref{eq:MBL-hamil} are taken from the reference so $U=24.4t$, the total particle number is 133 and we set $L=32$ with $N_r = 95$. In Fig.~\ref{fig:trap_results}$(a)$ we show the gap ratio of the QP spectrum related to a mean-field ground state with a central density of one surrounded by a condensate ring, contrary to the experiment which used a purely Mott-type initial state. The considered QP states localize at roughly the same energy scale as in the experiment, which we quantify by an exponential fit of $r$ for the least localized states at $\omega_{\gamma}/t \approx 0.1$ [see Fig.~\ref{fig:trap_results}$(b)$] resulting in a finite-size transition at $\Delta_c^{(r)}/t = 7.9(1.5)$ \footnote{Extremely small values of $r$ at small $\Delta$ and $\omega$ in Fig.~\ref{eq:trap_Greens}$(a,b)$ result from nearly degenerate low energy pairs due to an approximate discrete rotational symmetry.}.

To get further insight we consider the scaling of connected single particle correlations as given by
$
G_c(\ell,\ell') \equiv \langle \langle \hat{b}^{\dagger}_{\ell} \hat{b}_{\ell'} \rangle_{\textrm{QP}} - \phi^*_{\ell} \phi_{\ell'}\rangle_d.
$
Here $\langle \cdot \rangle_{\textrm{QP}}$ is the QP ground state expectation value implicitly defined via $\beta_{\gamma} | \psi_{\textrm{QP}} \rangle = 0$ for all $\gamma$ \cite{Bissbort2014,Geissler2018}, thus best fulfilling the original approximation of neglected QP interactions [now added above]. We then consider the radial correlations of the four central sites averaged for each unique distance from the trap center [see Fig.~\ref{fig:trap_results}$(c)$]. Due to the vicinity to a localization transition and the inhomogeneous nature of the system we expect an interplay of algebraic and exponential correlations which we summarize in the fit function \cite{Note4}
\begin{align} \label{eq:trap_Greens}
G_c(d) = a_1 \exp(-\lambda d) + a_2 d^{-b}.
\end{align}
In Fig.~\ref{fig:trap_results}$(d)$ we show the various obtained inverse localization lengths $\lambda$ of these fits together with one standard deviation of the fitting error. Below a certain disorder strength we find no exponential contribution. A linear fit for all nonzero $\lambda$ yields the theoretical critical disorder strength $\Delta_c^{(\lambda)}/t = 6.4(6)$ comparing well to the experimental value $\Delta_c/t = 5.3(2)$, which, to our knowledge, is the first theoretical prediction. The different slope compared to experiment likely stems from the slightly different nature of the considered observables.
We note that the localization happens at a much smaller disorder strength than predicted for the unconfined system. This is most likely due to the trap enhanced variance of the local potential. \\

\section{Conclusion}\label{sec:conclusion}

In conclusion, we have performed a detailed analysis of the two dimensional BHM with Gaussian disorder at half filling by discussing gap ratios and fractal dimensions of generalized (beyond Bogoliubov) QP eigenstates. We find a strongly localized spectrum with at least one mobility edge separating a small fraction of delocalized non-interacting QP modes at low energies from high lying localized ones. For all converged results this critical line follows an exponential decay with disorder down to quasiparticle energies of order $0.1t$.
Finite-size scaling in the vicinity of these critical lines yields relevant critical exponents and parameters for a spectral transition characterized by a thermal-like gap ratio and weak multi-fractality. Furthermore, the MEs are strongly affected by the structure of QP bands in the clean system. 
Our method predicts a scaling of correlations almost identical to that observed in experiment \cite{Choi2016} and the finite-size transition point without requiring any empirical fit parameter.

As we show in this work, the FOE is a very promising tool for the analysis of extended systems with strong correlations, which can also be used to clarify the interplay between MBL and the BG \cite{Geissler2020}. As the FOE can easily be extended to the time domain, it furthermore opens up an exciting direction of future research into disorder-driven dynamical effects.

\begin{acknowledgments}
A.G. would like to thank A. R. Abadal, C. Gro{\ss}, L. Rademaker and J. Schachenmayer for insightful discussions. Support by the Leopoldina Fellowship Programme of the German National Academy of Sciences Leopoldina grant no. LPDS 2018-14, the ANR ERA-NET QuantERA - Projet RouTe (ANR-18-QUAN-0005-01) and the High Performance Computing center of the University of Strasbourg, providing access to computing resources and scientific support, is gratefully acknowledged. Part of the computing resources were funded by the Equipex Equip@Meso project (Programme Investissements d'Avenir) and the CPER Alsacalcul/Big Data. G.P. is further supported by USIAS in Strasbourg and the Institut Universitaire de France (IUF).
\end{acknowledgments}

\bibliography{MBLlib}

%%%%%%%% Supplemental Material
\pagebreak
%\appendix
\onecolumngrid
\bigskip
\pagebreak
\begin{center}
\textbf{\large Supplemental Material for\\ ``Mobility edge of the two dimensional Bose-Hubbard model''}
\end{center}

%%%%%%%% reset counters
%\setcounter{equation}{0}
\setcounter{figure}{0}
\setcounter{section}{0}
%\setcounter{table}{0}
%\setcounter{page}{1}
%\makeatletter
\renewcommand{\theequation}{S\arabic{equation}}
\renewcommand{\thefigure}{S\arabic{figure}}

The numerical results presented in the main text are the result of an extensive finite-size scaling analysis of the quasiparticle (QP) spectrum for a disordered Bose-Hubbard model obtained using the fluctuation operator expansion method \cite{Bissbort2012,Bissbort2014,Geissler2018T,Geissler2018}. In the main text we also discuss the scaling of single particle correlations in an experimentally relevant system with harmonic confinement. Here, we provide further details on the finite-size scaling procedure and fitting of the correlations in the trapped system. In Sec.~\ref{sec:ME_scaling} we give an in-depth discussion of the procedure to obtain the finite-size scaling collapse of the level spacing ratios and the fractal dimensions of the quasiparticle wave functions in terms of the mean relative variances as a measure for the goodness of the collapse. We also comment on the limits of our numerical calculations due to the necessity of a basis truncation. Subsequently, in Sec.~\ref{sec:MF-BG} we use a very similar measure to quantify the presence of a structural ground state phase transition in the inhomogeneous mean-field ground state, which the FOE is based upon, consistent with earlier predictions for a zero-energy superfluid to Bose-glass transition \cite{Fisher1989,Makivic1993,Zhang1995,Priyadarshee2006,AlvarezZuniga2015}. We furthermore discuss possible scenarios for the decay of correlations of the QP ground state of the disordered Bose-Hubbard model in a harmonic trap in Sec.~\ref{sec:Experiment}, which we relate to experimental results in the main text. Finally, in Sec.~\ref{sec:entanglement} we point out the occurrence of many-body entanglement in the FOE, as its presence is highly relevant to the physics of many-body localization.

\section{Finite-size scaling at the Mobility Edge (ME)}\label{sec:ME_scaling}

First, we give a systematic finite-size scaling analysis of the \ac{QP} spectra revealing the mobility edge in the \ac{QP} fluctuations at given $U$ and $W$. For the two considered observables we obtain mutually consistent scaling exponents, which are also consistent with the Harris criterion \cite{Harris1974,Chayes1986,Vojta2014} implying that the critical point is not destabilized by Griffiths singularities. In general, second order phase transitions can be characterized by a generic algebraic scaling, which, for the two considered observables gap ratio $r$ and fractal dimension $D$ in the vicinity of the mobility edge, takes the form

\begin{align}
r_{L,W}(\omega) &= \tilde{r}_W \left( \left[\omega-\omega_c(W)\right] L^{1/\nu} \right), \label{eq:collapse_r_generic} \\
D_{L,W}(\omega) - D_c &= L^{-\beta/\nu} \tilde{D}_W \left( \left[\omega-\omega_c(W)\right] L^{1/\nu} \right). \label{eq:collapse_D_generic}
\end{align}
Here, $\tilde{r}_W(\cdot)$ and $\tilde{D}_W(\cdot)$ are the scaling functions, the index $W$ signifying a weak dependence on the disorder, while $L^{1/\nu}$ is the rescaled length scale. The universal scaling exponents $\nu$ and $\beta$ are to be determined numerically in combination with the ME critical energies $\omega_c(W)$ and the critical fractal dimension $D_c$. In the following we give a detailed discussion of the finite-size scaling collapse, which is performed at $U/t = 20$, $W/t\in[3,14]$ and $\omega_{\gamma}/t \in [0.710, 6.307]$ for $L \in \mathcal{L} = \lbrace 10,20,24,32,40 \rbrace$ with corresponding numbers of realizations $N_r \in \lbrace 480,240,240,95,48 \rbrace$. The range of QP energies and disorder strengths is chosen such as to cover as much as possible of the low energy ME between the lowest finite-size resolvable modes and the first band edge. We perform a binning for each $L$ and $W$ data set given by 30 equally spaced energies $\omega$ in the given interval. Each bin contains $N_b$ disorder averaged data points closest in mean QP energy $\omega_{\gamma}$ to each energy $\omega$, with $N_b \in \lbrace 4,8,12,20,32 \rbrace$ dependent on system size (chosen such that on average there is a slight overlap between the bins).

\subsection{Scaling analysis for the gap ratio \label{sec:scaling_gap}}

\begin{figure}[t]
 \centering
 %\captionsetup{width=0.95\textwidth}
 \includegraphics[width=0.49\columnwidth]{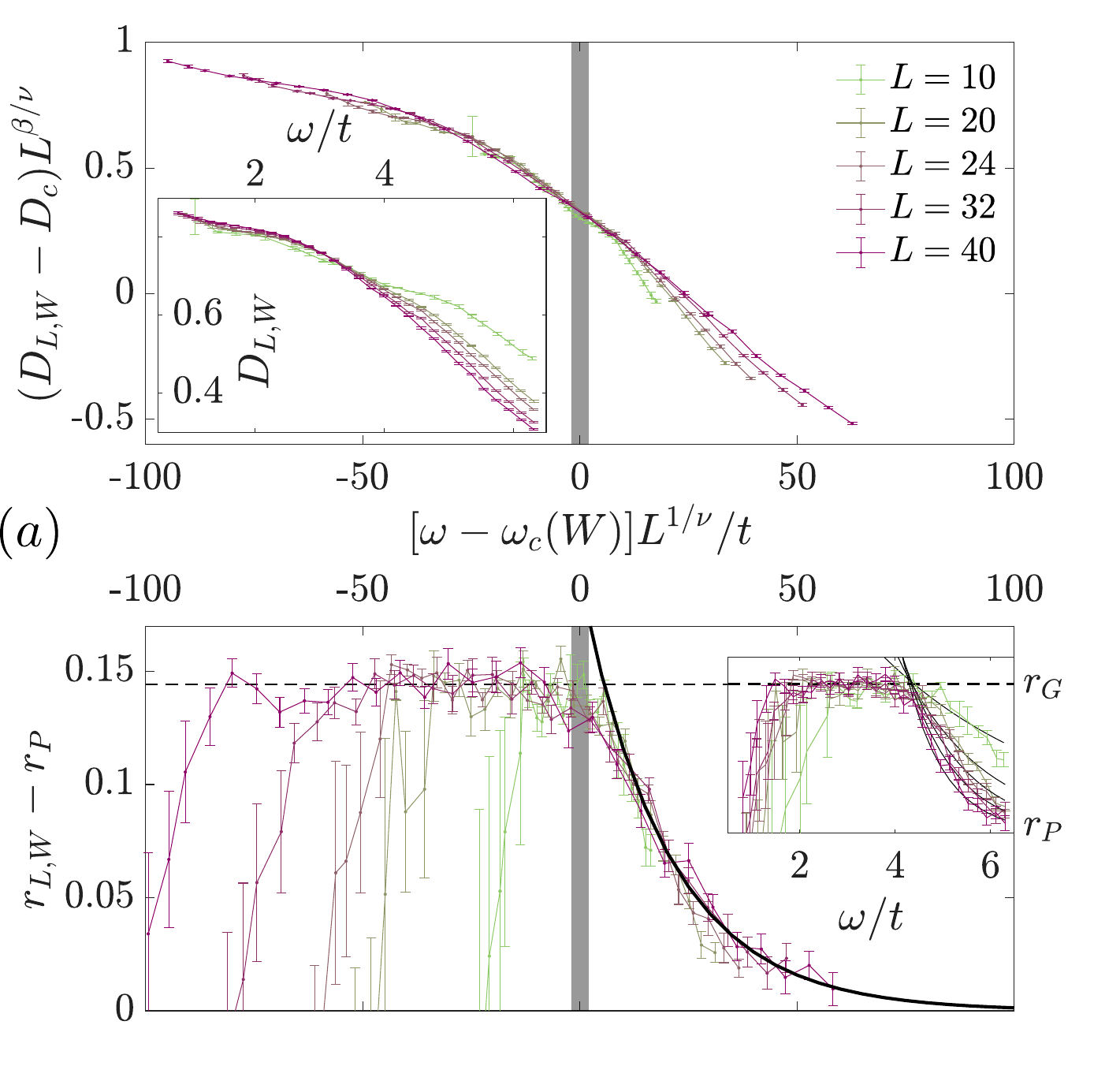}
 \includegraphics[width=0.49\columnwidth]{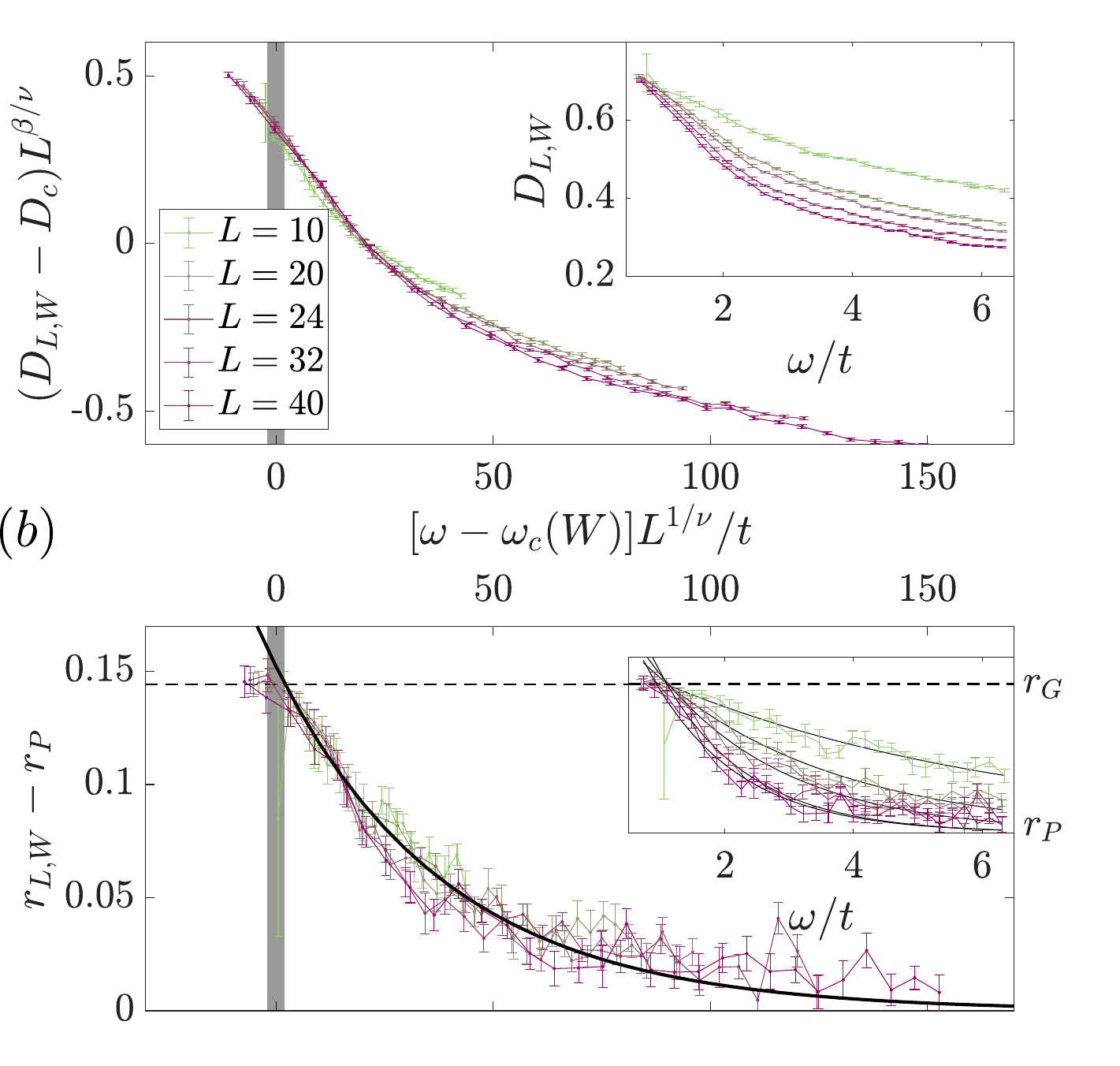}
 \caption{Exemplary scaling collapse of $D$ \textit{top} and $r$ \textit{bottom}. For the finite-size scaling every data set is binned for 30 equal spaced energies as described in the text, while $W/t \in \left\{4,11\right\}$ for $(a)$ and $(b)$, respectively. Data within the grey shaded regions is used to determine $\langle \tilde{D}_W(0) \rangle_{W}$ \textit{(top)} and $r_c$ \textit{(bottom)}. The dashed horizontal line in the \textit{bottom} row marks $r_{G}-r_P$ while the solid line is an exponential fit as guide for the eye. Insets in $(a)$ and $(b)$ show unscaled data while the black exponential fits of $r$ in the \textit{bottom} row are used to determine the ME $\omega_c(W)$.
 }
 \label{fig:more_r_Dmf_collapse}
\end{figure}

For the gap ratio $r$ the scaling ansatz \eqref{eq:collapse_r_generic}  involves a single scaling exponent in addition to the critical energies $\omega_c(W)$. In order to find the latter we note that $r$ as a function of the QP energy at given $W$ and for any system size $L$ always starts to decay from the thermal value $r_G$ at a common QP energy. The tails of this decay almost exactly follow an exponential, as visible in plots of the unscaled data [see main text and insets in Fig.~\ref{fig:more_r_Dmf_collapse}($b$)]. Such a behavior is consistent with a phase transition which in this case we expect to be a transition from delocalized to localized states in the QP spectrum. Thus the universality of phase transitions implies that exponential fits to the mentioned tails should cross in a single point at a certain QP energy, the critical energy corresponding to the ME. Thus we can obtain a remarkably good estimate of $\omega_c(W)$ for all $W/t \in \left[3,13 \right]$ considering the noise of the sampled $r$ data. Still, we also use differences in the decay constants of these exponential fits as weights in the averaging over the crossing points so the mean is not affected by large outliers due to nearby system sizes having very similar decay constants in their exponential fits (see for example inset Fig.~\ref{fig:more_r_Dmf_collapse}($b$, \textit{right})).  

Next we define a model independent measure for the goodness of the remaining single parameter scaling collapse involving only the parameter $1/\nu$. For this purpose it is beneficial to define the rescaled energies

\begin{align}
\bar{\omega}_{L,W} &= \left[\omega_{L,W}-\omega_c(W)\right] L^{1/\nu},
\end{align}
where the indices $L$ and $W$ are used to identify the individual binned data sets $r_{L,W}$ at all corresponding energies $\omega_{L,W}$. From these we define the interpolated functions $r_{L,W}(\bar{\omega})$. Then, the best choice of the scaling exponent $1/\nu$ minimizes the following mean relative variances,

%\begin{widetext}
\begin{align} 
\chi^{(W)}_r = \sum_{L \neq L'} \sum_{\bar{\omega}>0} &  \frac{\left({r}_{L,W} (\bar{\omega})  - {r}_{L',W} (\bar{\omega})\right)^2}{2 \left[ \sigma{r}^2_{L,W} (\bar{\omega}) + \sigma{r}^2_{L',W} (\bar{\omega})\right]}    \frac{r_{L,W}(\bar{\omega})}{\sigma r_{L,W}(\bar{\omega})} \frac{r_{L',W}(\bar{\omega})}{\sigma r_{L',W}(\bar{\omega})}  \frac{1}{\bar{C}_r^{(W)}}, \label{eq:alg_r_mean_rel_dev} 
\end{align}
%\end{widetext}
where $\sigma r_{L,W}$ are the standard errors of the mean determined from the binned data, while the normalization constants $\bar{C}_r^{(W)}$ are given by the total sum of all inverse relative covariances, used as weights to account for the noise,

\begin{align}
\bar{C}_r^{(W)} = \sum_{L \neq L'} \sum_{\bar{\omega}>0} \frac{r_{L,W}(\bar{\omega})}{\sigma r_{L,W}(\bar{\omega})} \frac{r_{L',W}(\bar{\omega})}{\sigma r_{L',W}(\bar{\omega})}.
\end{align}
Ideally, if all data points collapse onto the unknown scaling function within their error bars, this measure should be of order 1 or less. Note that we only consider $\bar{\omega}>0$ due the systematic finite-size effects resulting from symmetry related level bunching for low energy QP modes which are near plane wave excitations at weak disorder. Otherwise, we find that the finite-size scaling barely affects the delocalized states as $r_c \approx r_G$ implying a near-thermal critical behavior.

Due to the observed weak dependence of the scaling function $\tilde{r}_W$ on $W$, visible in the much slower decay of the collapsed data in Fig.~\ref{fig:more_r_Dmf_collapse}$(b)$ at large disorder, each disorder value $W$ is treated independently in the collapse of the binned gap ratio data quantified by \eqref{eq:alg_r_mean_rel_dev}. To approximate the error of the scaling collapse we sample over the results for all $W/t \in \left[ 3,13 \right]$ which we write as $\langle \cdot \rangle_W$. For the mean relative variances and its standard deviation we find $\langle \chi^{(W)}_r \rangle_W = 0.79(28)$. Regarding the universal exponent, the described minimization results in $1/\nu = 0.91(4)$. Sampling all collapsed data sets in the vicinity $|\bar{\omega}_{L,W}|/t < 2$ [marked by the vertical dark gray regions in Figs.~\ref{fig:more_r_Dmf_collapse}($b$)], chosen such that the value and its standard deviation are converged, further yields the critical gap ratio $r_c = \langle {r}_{L,W}(\bar{\omega} = 0) \rangle_{L,W} = 0.527(3)$ with $\langle \cdot \rangle_{L,W}$ the average over considered disorder values $W$ and system sizes $L$.

\subsection{Scaling analysis for the fractal dimension (q=2) \label{sec:scaling_Dmf}}

\begin{figure}[t]
 \centering
 %\captionsetup{width=0.95\textwidth}
 \includegraphics[width=0.32\columnwidth]{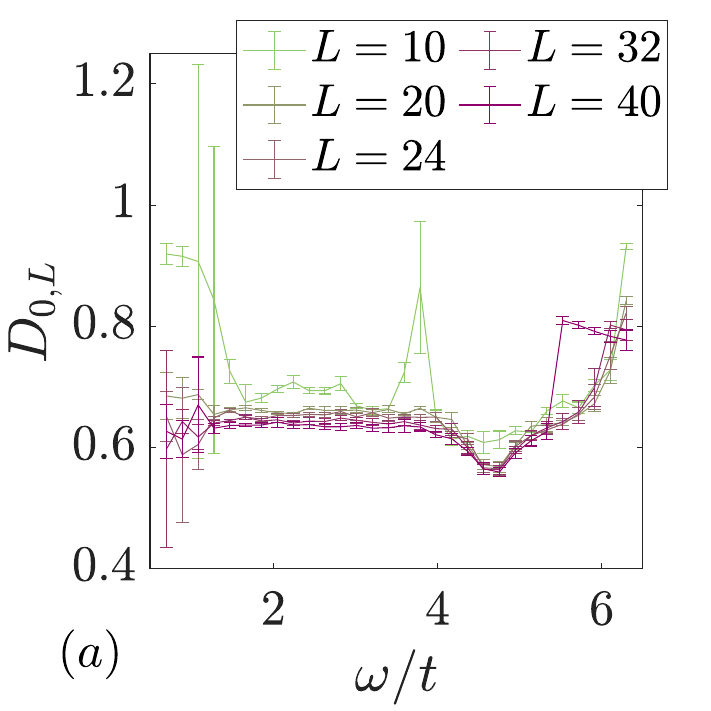}
 \includegraphics[width=0.32\columnwidth]{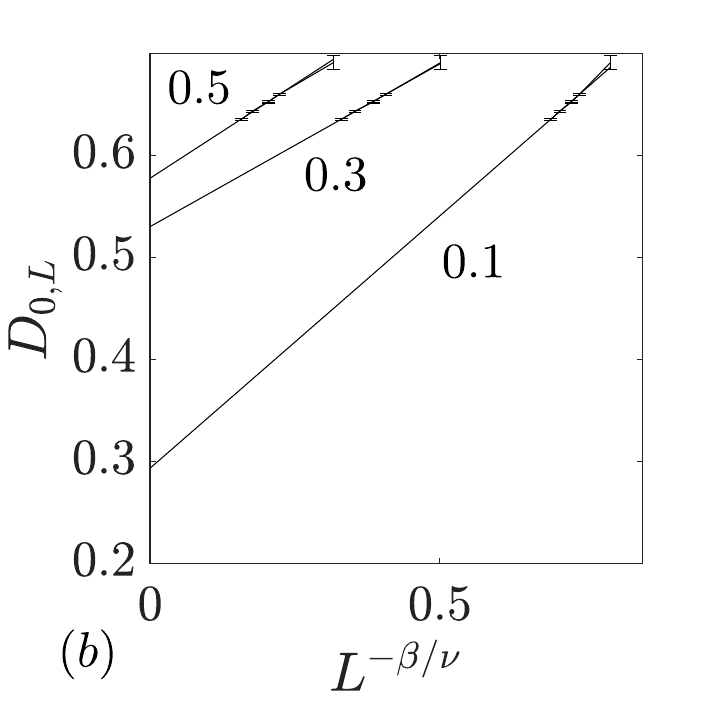}
 \includegraphics[width=0.32\columnwidth]{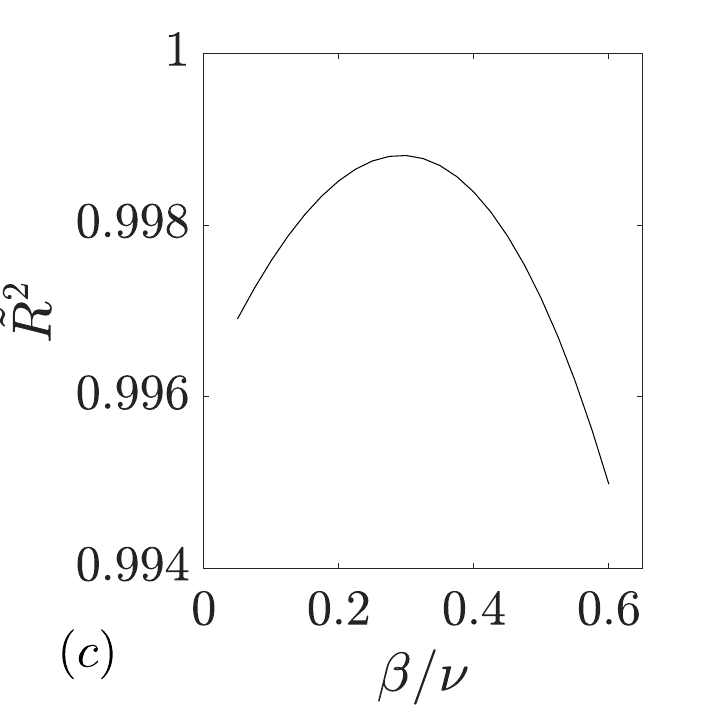}
 \caption{Finite-size scaling collapse of the inflection points of $D_{L,\omega}(W)$. $(a)$ Overview of the inflection points $D_{0,L}(\omega)$ as estimated from a fit of the binned data to $\tanh (\cdot)$. Exemplary best fit scaling collapses~\eqref{eq:collapse_D_inflection} of the inflection points sampled over $\omega / t \in \left[ 1.4, 3.7 \right]$ are shown in $(b)$. These fits use $\beta/\nu = \{ 0.1, 0.3, 0.5 \}$ as marked for each line, respectively. For a range of $\beta/\nu \in [0.05 , 0.6]$ $(c)$ shows the adjusted coefficient of determination $\tilde{R}^2$ for these fits.
 }
 \label{fig:inflection_perp}
\end{figure}

Finding the proper finite-size scaling of the fractal dimension is much harder compared to the gap ratio as Eq.~\eqref{eq:collapse_D_generic} has twice the number of scaling exponents and parameters at any given disorder $W$. To find a unique prediction we determine the parameters step by step via independent means. Firstly, we note that for fixed QP energies and as a function of disorder $W$ the fractal dimension always has a sigmoid shape. At the same time finite-size scaling tells us that the inflection points $W_{L,\omega}$ of such a series of sigmoid functions have to collapse onto a single point $\bar{W}_0(\omega)$ of the scaling function. Correspondingly, for any of these fixed QP energies $\omega$, where $\bar{W}_{0}(\omega) = \left[W_{L,\omega}-W_c(\omega)\right] L^{1/\nu}$ is the scaled disorder, in analogy to Eq.~\eqref{eq:collapse_D_generic} one also gets

\begin{align}
D_{0,L} \equiv D_{L}(W_{0,L}) = \frac{\tilde{D} \left( \bar{W}_0 \right)}{L^{\beta/\nu}} + D_c. \label{eq:collapse_D_inflection}
\end{align}

\begin{figure}[b]
 \centering
 %\captionsetup{width=0.95\textwidth}
 \includegraphics[width=0.65\columnwidth]{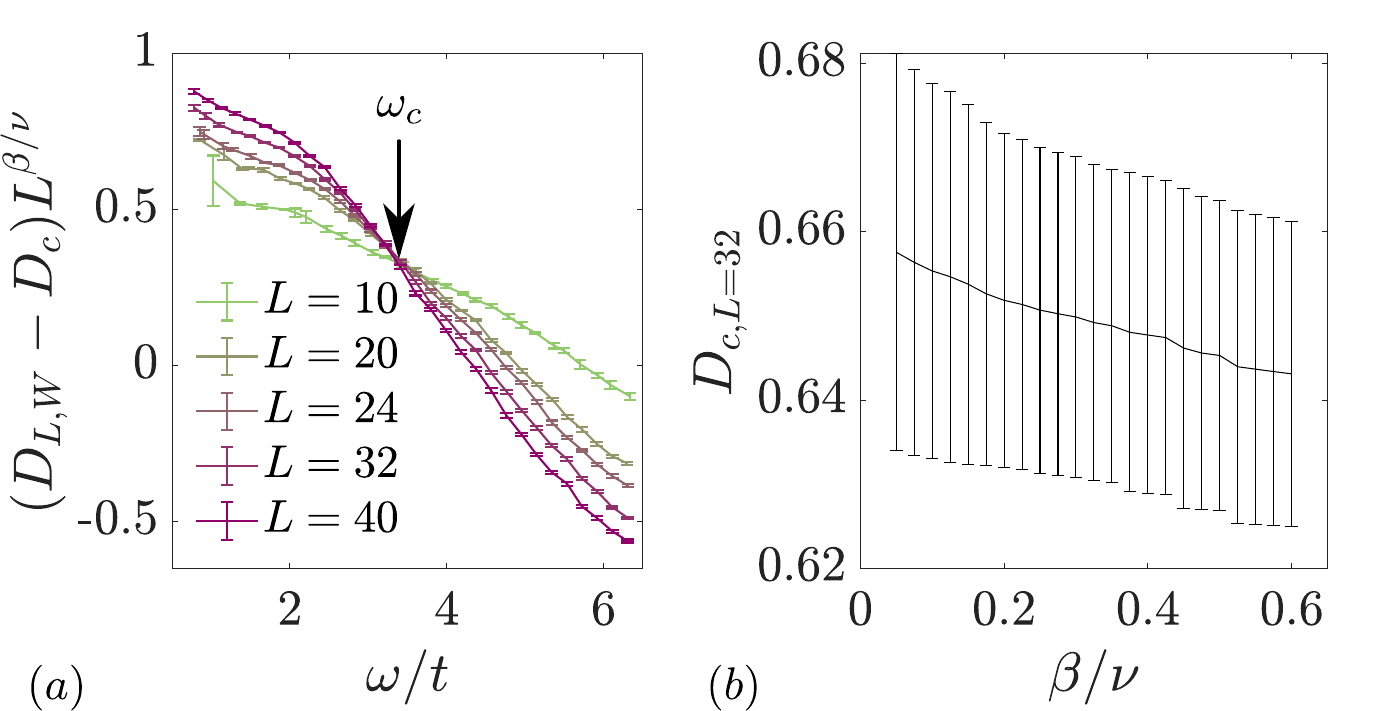}
 \caption{Semi-scaling collapse at the critical ME $\omega_c(W)$. ($a$): Finite-size scaled fractal dimension $D$ of the binned QP spectrum in the energy interval considered for the scaling with $U/t=20$, $W/t=5$, $\beta/\nu = 0.26$ and $D_c=0.51$. System sizes are given in the legend, while the black arrow signals the mutual crossing point of the data for all system sizes. ($b$): Finite-size critical dimension as a function of the scaling exponent $\beta/\nu$, determined via the crossing points and sampled over all disorder realizations with $W/t \in [3,13]$. The errors are given by one standard deviation.
 }
 \label{fig:ME_from_semiscaled_Dmf}
\end{figure}

As we know the fractal dimension only for a limited set of disorder values we use a fit to a generic sigmoid function [$\tanh (\cdot)$] to determine the inflection points and thus $D_{0,L}$ as a function of $L$. A few examples of such fits are shown in the insets of Figs.~\ref{fig:r_Dmf_collapse_perpendicular}($a$). In order to obtain a sufficiently good estimate of an inflection point it cannot be too close to the edges of the data range or even outside. We find the most consistent estimates to be in the range of energies $\omega/t \in \left[ 1.4, 3.7 \right]$ (see Fig.~\ref{fig:inflection_perp}($a$)). As all $D_{0,L}$ within this range are consistent with each other we can take an average over these QP energies. The resulting values we then fit by Eq.~\eqref{eq:collapse_D_inflection} where we consider $\tilde{D} \left( \bar{W}_0 \right)$ as an unknown parameter, such that this fit is slightly over-determined. Thus we take $\beta/\nu$ as a fixed parameter of these fits to obtain sets of best $D_c$ and $\tilde{D} \left( \bar{W}_0 \right)$ as functions of $\beta/\nu$. We find a wide range of $\beta/\nu$ between 0.1 and 0.5 giving very similar quality of fits [compare Fig.~\ref{fig:inflection_perp}($b$)], as quantified by the adjusted coefficient of determination $\tilde{R}^2$ shown in Figs.~\ref{fig:inflection_perp}($c$).

In a next step we determine the critical QP energies $\omega_c$ from the mutual crossing points of the scaled fractal dimension as a function of QP energies for fixed disorder values, using $D_c(\beta/\nu)$ from the previous step. Figure~\ref{fig:ME_from_semiscaled_Dmf}($a$) shows and example using $\beta/\nu=0.26$ with $D_c = 0.51$. The collection of all these crossing points yields the ME $\omega_c(W)$ as well as the finite-size critical dimension $D_{L,W}(\omega_c(W)) = \frac{\tilde{D}_W(0)}{L^{\beta/\nu}} + D_c$. Prior knowledge of the scaling function $\tilde{D}_W(0)$ could also be used to determine the critical ME for any single finite-size system without having to perform a full finite-size collapse. But as the scaling function $\tilde{D}_W(\bar{\omega}_c)$ has a weak disorder dependence, also $D_{L,W}(\omega_c(W))$ varies slightly along the ME. As the scaling function is non-universal, we approximate the error of the disorder averaged finite-size critical dimension $D_{c,L} = \langle D_{L,W}(\omega_c(W)) \rangle_W$, for $W/t \in [3, 13]$, by its standard deviation to account for these variations. As shown in Fig.~\ref{fig:ME_from_semiscaled_Dmf}($b$) for $L=32$ the finite-size critical dimension $D_{c,L}$ and thus also the predicted ME barely depends on the scaling exponent $\beta/\nu$ within the optimal region [see Fig.~\ref{fig:inflection_perp}($c$)].

Finally, keeping all parameters including the ME at all considered disorder values except $1/\nu$ fixed, we perform the final collapse by minimizing the following mean relative variances

%\begin{widetext}
\begin{align} 
\chi^{(W)}_D = \sum_{L \neq L'} \sum_{\bar{\omega}}   \frac{\left(\tilde{D}_{L,W} (\bar{\omega})  - \tilde{D}_{L',W} (\bar{\omega})\right)^2}{2 \left[ \sigma\tilde{D}^2_{L,W} (\bar{\omega}) + \sigma\tilde{D}^2_{L',W} (\bar{\omega})\right]} \frac{1}{\bar{C}_D}, \label{eq:alg_D_mean_rel_dev} 
\end{align}
%\end{widetext}
where $\sigma D_{L,W}$ are the standard errors of the mean determined from the binned data, while the normalization constant $\bar{C}_D$ is given by the total number of terms, $\bar{C}_D = \sum_{L \neq L'} \sum_{\bar{\omega}} 1$. As stated before, for an ideal collapse such a measure should be of order one.

\begin{figure}[b]
 \centering
 %\captionsetup{width=0.95\textwidth}
 \includegraphics[width=0.32\columnwidth]{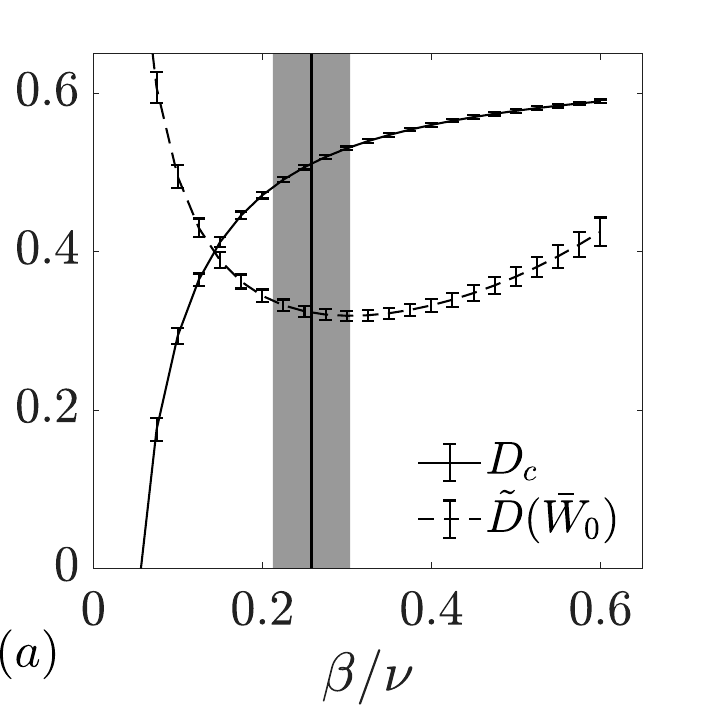}
 \includegraphics[width=0.32\columnwidth]{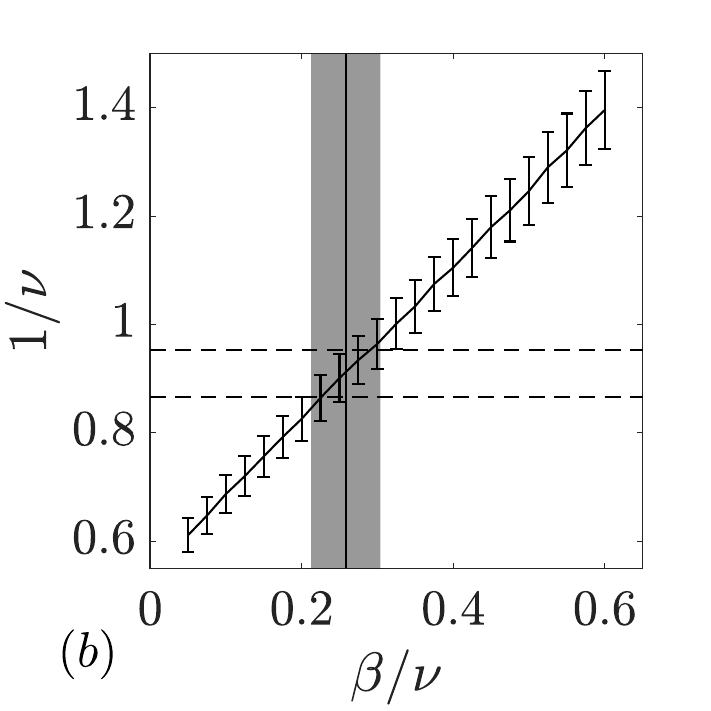}
 \includegraphics[width=0.32\columnwidth]{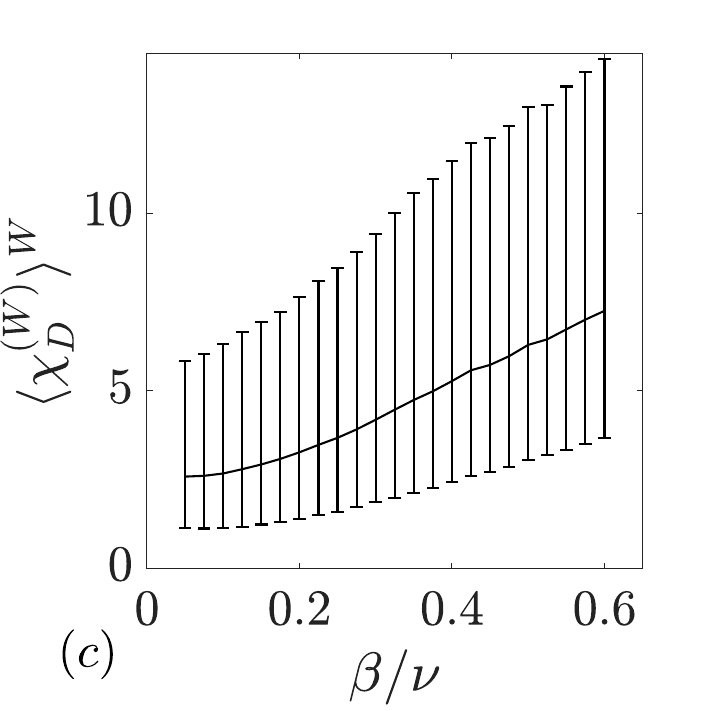}
 \caption{Finite-size scaling collapse of the fractal dimension. $(a)$ Best fits for the fixed QP energy inflection points $D_L(W_{0,L})$ to the scaling relation ~\eqref{eq:collapse_D_inflection} yield $D_c$ and $\tilde{D}(\bar{W}_0)$ as functions of $\beta/\nu$ with errors representing the one standard deviation confidence interval of the best fit. $(b)$ Minimization of~\eqref{eq:alg_D_mean_rel_dev} for each $W$ with respect to $1/\nu$ reveals the linear relation of the best $1/\nu$ and $\beta/\nu$ Errors are standard errors of the mean resulting from a sampling of the best $1/\nu$ over the considered disorder values $W/t \in [3 ,13]$. Similarly, in $(c)$ we show the typical mean relative variances $\langle \chi^{(W)}_D \rangle^W$ as measure of the quality of the collapse together with its standard deviation.
 }
 \label{fig:Dmf_full_collapse}
\end{figure}

Now, the minimization of Eq.~\eqref{eq:alg_D_mean_rel_dev} is performed for each considered disorder value $W/t \in [3,13]$ and for various $\beta/\nu \in [0,0.6]$. Each $\beta/\nu$ comes with a ME $\omega_c(W)$ and critical dimension $D_c$ [see Fig.~\ref{fig:Dmf_full_collapse}($a$)]. Averaging the resulting best $1/\nu$ over the considered disorder data sets reveals a nearly linear relation between $\beta/\nu$ and $1/\nu$ of the best collapses. Requiring that $1/\nu$ has to be identical to the value obtained for the collapse of the gap ratio data in the preceding section ($1/\nu = 0.91(4)$) yields the prediction for the scaling exponent $\beta/\nu = 0.26(5)$, as shown in Fig.~\ref{fig:Dmf_full_collapse}($b$). We note that these scaling exponents are consistent with those of the one-dimensional directed percolation universality class \cite{Jensen1999,Wang2013}. Due to some rare systematic finite-size effects for small system sizes and weak disorder (compare finite-size prediction of the mobility edge for $L=10$ shown in Fig.~\ref{fig:finite-size-MEs}), we quantify the quality of these best collapses by considering the typical value of Eq.~\eqref{eq:alg_D_mean_rel_dev}, which we obtain using the definition $\langle \cdot \rangle^W = \exp\left[ \langle \ln(\cdot) \rangle_W \right]$ and its standard deviation, see Fig.~\ref{fig:Dmf_full_collapse}($c$). As a result of some finite-size corrections, especially for the smallest system sizes (compare examples in Fig.~\ref{fig:more_r_Dmf_collapse} and $L=10$ finite-size prediction of the ME in Fig.~\ref{fig:finite-size-MEs}), this measure is slightly larger then one due to some large deviations of the full finite-size collapse appearing far from the critical point [note especially the very small errors in the fractal dimension at very low QP energies in Fig.~\ref{fig:more_r_Dmf_collapse}($a$)].

\subsection{Further scaling observations and remarks}

\begin{figure}[h]
 \centering
 %\captionsetup{width=0.95\textwidth}
 \includegraphics[width=0.32\columnwidth]{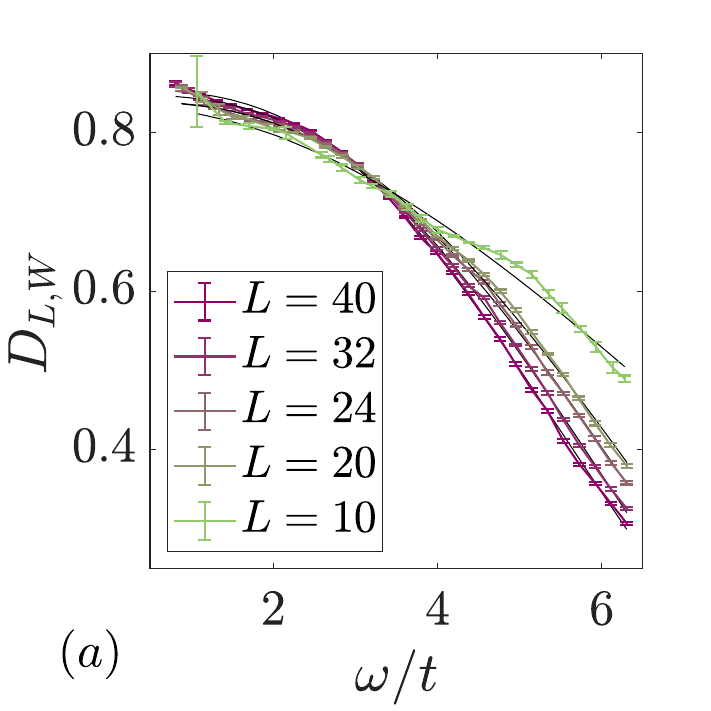}
 \includegraphics[width=0.32\columnwidth]{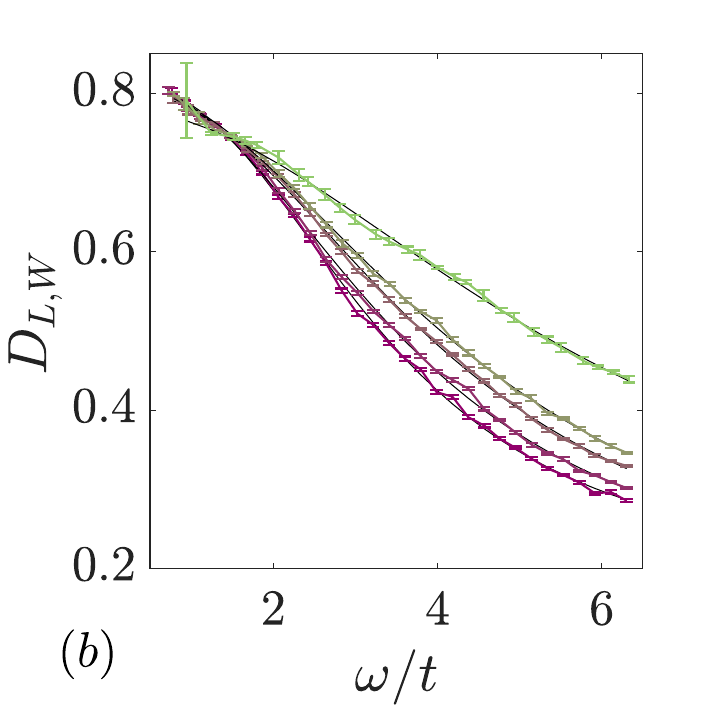}
 \includegraphics[width=0.32\columnwidth]{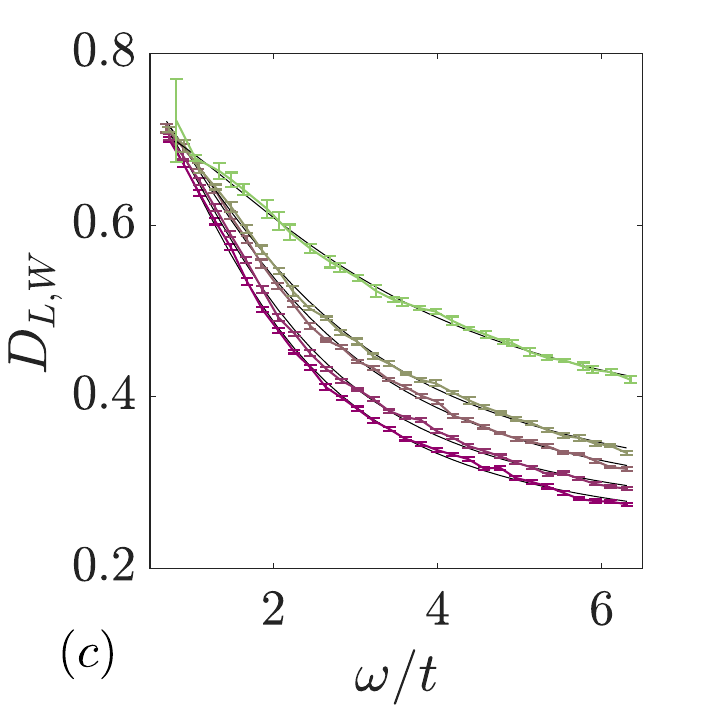}
 \caption{Spectra of the fractal dimension $D$ of the binned QP spectrum for $\omega/t \in [0.710, 6.307]$ and various system sizes, given in the legend $(a)$. Shown are examples for $W/t \in \{4,7,11\}$ in $(a)$, $(b)$ and $(c)$, respectively. The black lines are best fits to the empirical ansatz~\eqref{eq:modified_sigmoid}.
 }
 \label{fig:Dmf_inflection_fixW}
\end{figure}

\begin{figure}[t]
 \centering
 %\captionsetup{width=0.95\textwidth}
 \includegraphics[width=0.32\columnwidth]{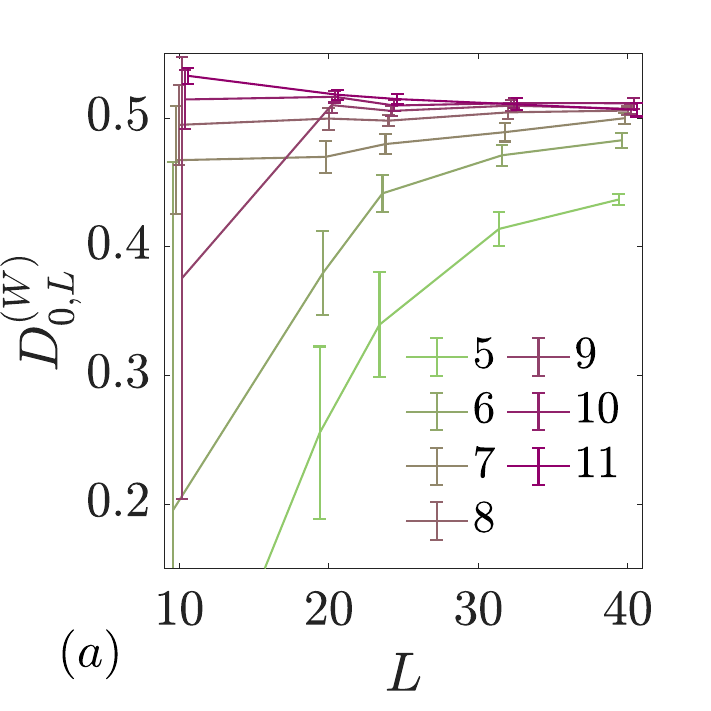}
 \includegraphics[width=0.32\columnwidth]{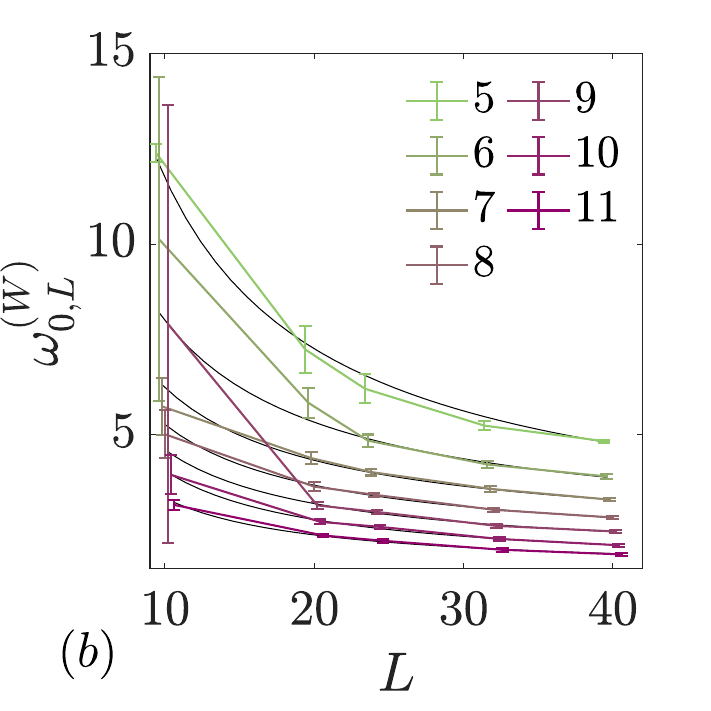}
 \includegraphics[width=0.32\columnwidth]{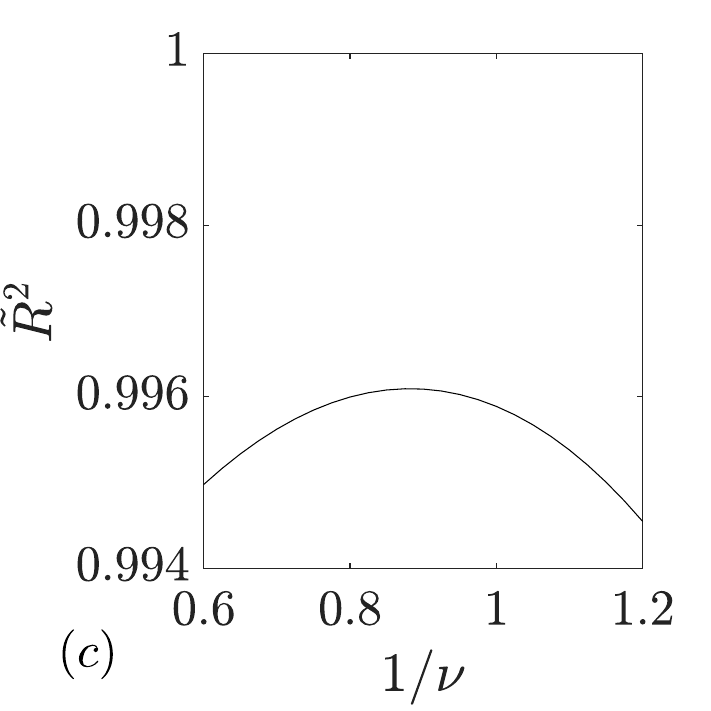}
 \caption{Finite-size scaling of the inflection points [$D^{(W)}_{0,L}$ in $(a)$ and $\omega^{(W)}_{0,L}$ in $(b)$] of the fractal dimension as function of the QP energy at various disorder values $W/t$ given in the legend. For better readability the data sets are slightly offset in $L$ relative to each other. Black lines in $(b)$ are best fits to eq.~\ref{eq:collapse_om_inflection} for fixed $1/\nu = 0.91$. For a range of fixed values $1/\nu \in [ 0.6 1.2 ]$ $(c)$ shows the adjusted coefficient of determination $\tilde{R}^2$ for the fits of the combined data sets with $W/t \in [7,11]$ as described in the text.
 }
 \label{fig:Dmf_inflection_fixW_collapse}
\end{figure}

The idea to follow the finite-size scaling of special points of the scaling function can also be applied to the fractal dimension of the QP states $D_{L,W}(\omega)$ as a function of energy $\omega$ for fixed disorder $W$ and various system sizes $L$. But in this case the previously used ansatz of using a $\tanh(\cdot)$-shaped function to determine the inflection points does not work due to the strong asymmetry of $D_{L,W}(\omega)$. Instead, motivated by the empirically obtained shape of the ME $\omega_c(W) = \omega_0 \exp(-W/\Omega) $, we instead consider the fit function

\begin{align}
 f(\omega) = a \tanh\left[ b \left( \ln \omega^{(W)}_{0,L} - \ln \omega \right) \right] + D^{(W)}_{0,L}. \label{eq:modified_sigmoid}
\end{align} 
Via this fit we find the approximate inflection points $\left(\omega^{(W)}_{0,L},D^{(W)}_{0,L}\right)$. Similar to before, this procedure works best when the QP energy of the inflection point is sufficiently far from the considered energy limits, so for $W/t \in [5,11]$. Furthermore, at weak disorder and low QP energies we find some deviation from the empirical fit function (compare Figs.~\ref{fig:Dmf_inflection_fixW}). In Figs.~\ref{fig:Dmf_inflection_fixW_collapse}$(a,b)$ we show the scaling behavior of $D^{(W)}_{0,L}$ and $\omega^{(W)}_{0,L}$. Very clearly these data points imply $D^{(W)}_{0,L} \approx 0.5$ for $L \rightarrow \infty$, or more precisely $\langle D^{(W)}_{0,L=40} \rangle_W = 0.506(4)$ for $W/t \in [7,11]$ with the error given by the standard deviation. This is consistent with the value of the critical fractal dimension $D_c$ determined in the finite-size scaling above. Only for $W/t < 7$, where we expect the fit to be less reliable, do we observe apparent finite-size scaling behavior, but with relatively large uncertainties for the inflection points. On the other hand $\omega^{(W)}_{0,L}$ shows very consistent scaling behavior, such that we can attempt a scaling collapse similar to~Eq.\eqref{eq:collapse_D_inflection}:

\begin{align}
\omega^{(W)}_{0,L} = \frac{\bar{\omega}^{(W)}_{0,L}}{L^{1/\nu}} + \omega_c(W). \label{eq:collapse_om_inflection}
\end{align}
As $\omega_c(W)$ is the mobility edge and as such not a constant, we  combine the data sets $\omega_{0,L} = \langle \omega^{(W)}_{0,L}\rangle_W$ for $W/t \in [7,11]$ where we consider the inflection points to be most reliable. Via this sampling we reduce the noise in the combined data so we are best able to determine the optimal scaling exponent by fitting to the average $\omega_{0,L} = \bar{\omega}_{0,L} / L^{1/\nu} + \omega_c$ of~Eq.\eqref{eq:collapse_om_inflection} over the disorder for various values of $1/\nu$ and considering the adjusted coefficient of determination $\tilde{R}^2$ shown in Fig.~\ref{fig:Dmf_inflection_fixW_collapse}. While we find a wide range of parameters that give very similar values of $\tilde{R}^2$ the optimum is at about $1/\nu \approx 0.9$ consistent with our earlier finding. In Fig.~\ref{fig:Dmf_inflection_fixW_collapse} we also show fits of Eq.~\eqref{eq:collapse_D_inflection} with $1/\nu = 0.91$ which all fit the inflection point data within the respective errorbars.

\begin{figure}[h]
 \centering
 %\captionsetup{width=0.95\textwidth}
 \includegraphics[width=0.99\columnwidth]{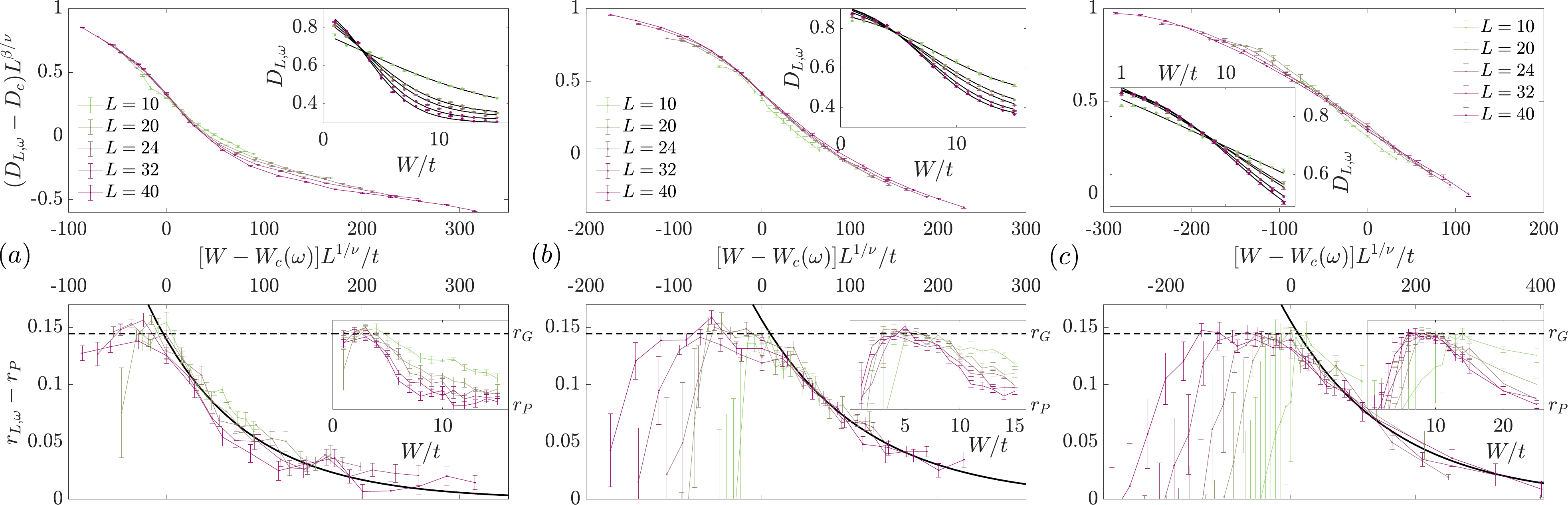}
 \caption{Exemplary scaling collapses of $D$ \textit{(top)} and $r$ \textit{(bottom)} as functions of disorder strength $W$ at fixed QP energies. Here we considered the binned data (with $N_b$ disorder averaged data points) centered at the energies $\omega/t = \{4.1,2.1,1.1\}$ corresponding to $\omega_c(W)$ at $W/t \in \{4,7,11\}$ for $(a)$, $(b)$ and $(c)$, respectively. Insets show unscaled data with black $\tanh(\cdot)$ fits for $D$ in the \textit{top} row used to determine the inflection points $D_{L}(W_{0,L})$. In the \textit{bottom} row the horizontal dashed line marks $r_{G}-r_P$ and the solid line is an exponential fit as guide for the eye. 
 }
 \label{fig:r_Dmf_collapse_perpendicular}
\end{figure}

As we have seen, the scaling parameters are consistent both for a finite-size scaling along the QP energies and fixed disorder, as well as for variable disorder and fixed QP energies. Therefore, for the sake of completeness we also show exemplary full collapses at fixed QP energies $\omega$ using $\omega_c(W_c)$ with $W_c/t \in {4,7,11}$ as the binning centers. The resulting scaling collapses are shown in Fig.~\ref{fig:r_Dmf_collapse_perpendicular} for the scaling exponents and parameters determined earlier. Indeed, these parameters result in remarkably good collapses for the different parts of the spectrum.

\begin{figure}[t]
 \centering
 %\captionsetup{width=0.95\textwidth}
 \includegraphics[width=0.99\columnwidth]{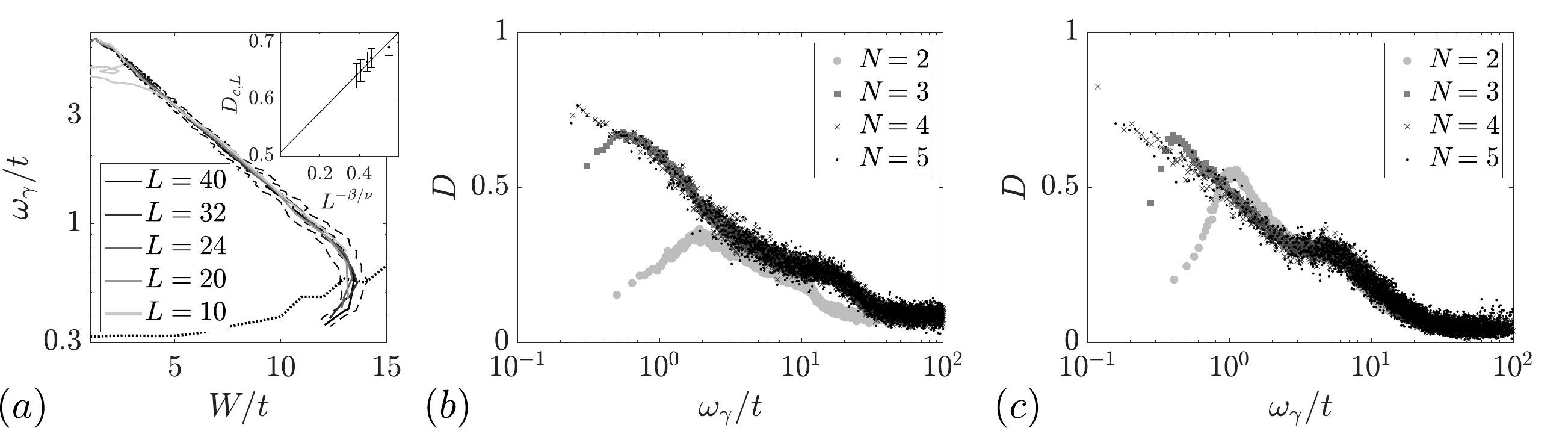}
 \caption{
$(a)$: Mobility edge for $U/t = 20$ as given by fractal dimension contour lines at $D_{c,L} = D_c + \tilde{D}(0)/L^{\beta/\nu}$ for the considered system sizes $L \in \left\{10,20,24,32,40 \right\}$ given in the legend with the data binned for every four consecutive energy levels. Each $D_{c,L} = \langle D_{L,W}(\omega_c(W)) \rangle_W$ shown in the inset is sampled along the mobility edge $\omega_c(W)$ for $W/t \in [3,13]$ [e.g. shown in Fig.~\ref{fig:ME_from_semiscaled_Dmf}($a$) for $W/t = 5$]. The errors of $D_{c,L}$ in the inset are one standard deviation for this sampling each, which is mostly dominated by systematic variations in the scaling functions $\tilde{D}_W(\bar{\omega})$. For the black line in the inset we use the average value $\tilde{D}(0) = \langle \tilde{D}_W(\bar{\omega}=0) \rangle_W$ and $\beta/\nu = 0.26$, as determined from the scaling collapse. The dashed lines in the main panel correspond to the lower and upper contour at $L=40$ of one SD in the value of $D_{c,L}$ (see inset), while the dotted line marks the position of the maximum of $D(\omega_{\gamma})$ for $L=40$ [as estimator for the truncation limit, see $(b)$] with energies binned for every four consecutive QP levels. $(b,c)$: Effect of the truncation of the number of mean-field eigenstates $N$ (see legend) considered in the FOE on the convergence of the fractal dimension of the QP eigenstates. Here, the system size is $L=32$ for all cases, while $(U/t,W/t) = (20,13)$ in $(b)$ and $(U/t,W/t) = (3,8)$ in $(c)$. Numerical simulations for each truncation $N \in \lbrace 2,3,4,5 \rbrace$ have been performed over $N_r \in \lbrace 95,95,20,10 \rbrace$ identically seeded disorder realizations, respectively.
 }
 \label{fig:finite-size-MEs}
\end{figure}

Finally, in the main text we also show the finite-size scaling prediction of the ME in the thermodynamic limit by considering the finite-size critical contour $D_{L,W}(\omega) = D_{c,L} = D_c + \tilde{D}(0)/L^{\beta/\nu}$ where $D_c$ is the critical fractal dimension and $\tilde{D}(0)$ is the value of the scaling function at the critical point. As can be seen in Fig.~\ref{fig:finite-size-MEs}, all finite size predictions of the ME match almost exactly except for some deviations for small system sizes. We note that the reentrant shape of the contour lines at strong disorder and low energies is a truncation artifact as we will discuss now.

We note that, especially at large energies as well as the low excitation energies relevant here, another scaling in addition to $L$ becomes important as well. This is related to the Bose statistics, namely its unbounded local Fock-bases of bosonic number states, which in numerical simulations is commonly truncated at some sufficiently large number $N_b$. Within the FOE method this truncation is realized on the level of considered mean-field Gutzwiller (eigen-)states at each site, $N$. We always use $N_b = 3 N$ to get a good approximation of the low energy part of the considered local operators in terms of the MF states. Furthermore, in the main part we consider the up to $N = 5$ lowest MF states at each site which is sufficient to resolve the ME down to $\omega_{\gamma} \approx 0.1t$, while for the energy and disorder window considered for the finite-size scaling (see main text and Sec.~\ref{sec:scaling_gap} and~\ref{sec:scaling_Dmf}) $N=3$ already is sufficient, as we will see in a moment. Only as the ME approaches the lowest resolved QP energy levels -- those following an unphysical reentrant shape (see Fig.~\ref{fig:finite-size-MEs}$(a)$) -- do we observe a pronounced dependence of the numerical results for the (multi-)fractal dimension $D$ of the QP states on the truncation $N>2$, as we show in Figs.~\ref{fig:finite-size-MEs}($b,c$) for weak and strong interaction $U/t \in \lbrace 3,20 \rbrace$, respectively. Comparing $N=3$ with $N>3$ one can see that in terms of $D$ the QP states are well converged at all QP energies down to those QP levels $\omega_{\gamma}$ for which $D(\omega_{\gamma})$ has its unphysical maximum at $N=3$. In Fig.~\ref{fig:finite-size-MEs}$(a)$ the dotted line represents the energy of this maximum as an approximate bound below which the QP states can not be considered converged in terms of the considered local basis truncation $N=3$.

\section{\label{sec:MF-BG}Characterization of the mean-field ground state phase transition}

It has long been established that the ground state of a disordered two-dimensional Bose-Hubbard model, as used in the main part, can exhibit the formation of a so-called Bose-glass phase \cite{Fisher1988,Fisher1989,Herbut1997,Soyler2011,AlvarezZuniga2015}. In this section we briefly comment on the relation of our results to these earlier predictions on the level of the fragmentation of the mean-field ground state. As we will show, such a transition for finite values of $U/t$ can be also determined by an in-homogeneous mean-field description which at the same time forms the basis of the FOE method used in the main text and above. Our results are consistent with a direct phase transition from a condensed superfluid to a Bose-glass in the ground state. For example, earlier works using a uniform disorder distribution  $[-W,W]$ (best compared to the full width at half maximum $\Delta = 2 \sqrt{2 \ln 2} W \approx 2.35 W$ of a Gaussian distribution with standard deviation $W$) have predicted this transition to be at $W_c/t \approx 5$ \cite{Fisher1989,Makivic1993,Zhang1995,Priyadarshee2006,AlvarezZuniga2015}  at half-filling and in the limit of infinite interaction $U/t \rightarrow \infty$, the so-called hard-core boson case.

To characterize the phase transition in the ground state of the disordered Bose-Hubbard model in two dimensions we consider the $q \rightarrow \infty$ limit of the multi-fractal dimensions as an order parameter. Applied to the distribution of the inhomogeneous mean-field condensate order parameter $\phi_{\ell}$ and sampled over the disorder realizations $\langle \cdot \rangle_d$ we define

\begin{align}
D_{\phi} = - \left\langle \log_L \left( \frac{\textrm{max}_{\ell}|\mathbf{\phi}_{\ell}|^2 }{ \sum_{\ell}^{L^2} |\mathbf{\phi}_{\ell}|^2} \right) \right\rangle_d.
\end{align}
As in the main part we evaluate this observable over a range of disorder strengths $W/t \in [1, 16]$, fixed interaction $U = 20t$ and for the same linear system sizes $L \in \mathcal{L} = \left\{ 10,20,24,32,40 \right\}$ while averaging over $N_d = 60$ disorder realizations each time. 

\begin{figure}[b]
 \centering
 %\captionsetup{width=0.95\textwidth}
 \includegraphics[width=0.99\columnwidth]{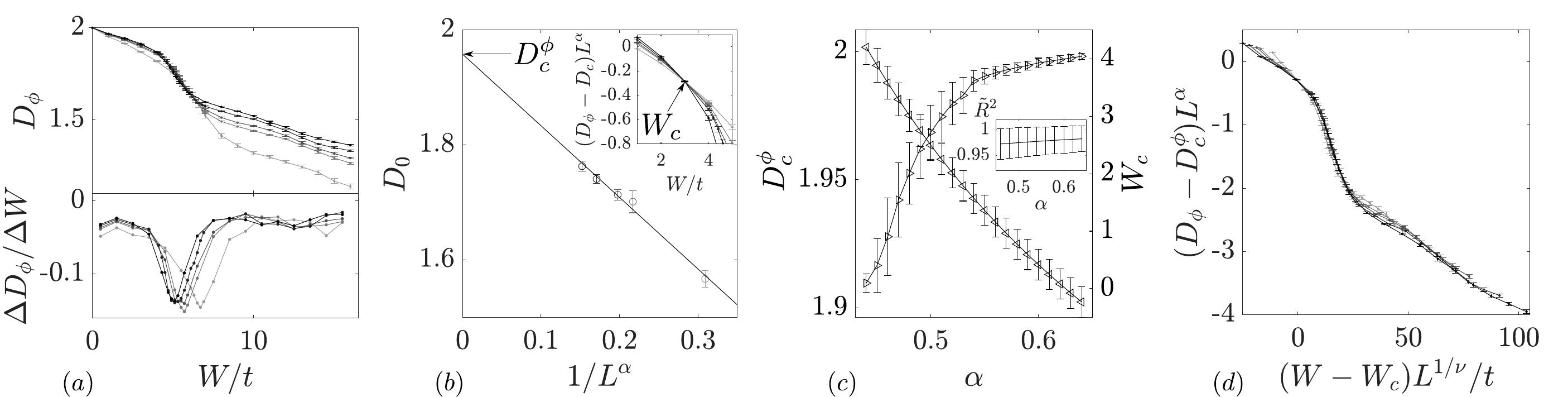}
 \caption{
Finite-scaling collapse of the structural transition of the MF condensate order parameter. $(a)$ depicts the fractal dimension $D_{\phi}$ of the MF parameter $\phi_{\ell}$ (\textit{top}) and its difference quotient (\textit{bottom}). Colors ranging from light grey to black correspond to the considered system sizes $L \in \left\{ 10,20,24,32,40 \right\}$, respectively. Same colors are used in $(b)$ showing the scaling collapse of the fractal dimension at the inflection point $D_0(L)$. The critical dimension $D_c$ is marked by a black arrow for the case $\alpha = 0.51$. Both parameters together give the partial collapse in the inset. The mutual crossing point of all data sets yields the critical disorder strength $W_c$ marked by a black arrow. $(c)$ Parameter $D_c^{\phi}$ for the best fits of $D_0(L)$ to the scaling ansatz~\eqref{eq:collapse_Dphi_inflection} for various values of $\alpha$ (left pointing triangles) and $W_c$ obtained from the crossing points of the partial scaled data sets (right pointing triangles). The adjusted coefficient of determination $\tilde{R}^2$ for the best fits is given as an inset. In $(d)$ the combined full collapse is shown for $\nu = 1.77$ and $\alpha = 0.51$. The corresponding critical point $(W_c,D_c)$ is taken from the relation shown in panel ($c$).
 }
 \label{fig:SF-BG-scaling_U20}
\end{figure}

In Fig.~\ref{fig:SF-BG-scaling_U20}($a$) we show $D_{\phi}$ and its difference quotient revealing  an inflection point $D_0(L) = D_{\phi}[W_0(L)]$ in the function $D_{\phi}(W)$, which experiences a finite-size scaling shift. We obtain $D_0(L)$ via parabola-fits to the minima in the difference quotient. Analogous to the mobility edge we consider a scaling ansatz of the form 
\begin{align}
D_{\phi,L}(W) - D^{\phi}_c = L^{-\alpha} \tilde{D} \left( \left[W-W_c\right] L^{1/\nu} \right). \label{eq:collapse_Dphi_generic}
\end{align}
For the scaling collapse of the inflection points $D_0(L)$ onto the inflection point of the scaling function $\tilde{D}(\bar{W})$, where $\bar{W} = \left[W(L)-W_c\right] L^{1/\nu}$ is the rescaled disorder, we thus expect
\begin{align}
D_0(L) = \frac{\tilde{D}(\bar{W}_0)}{L^{\alpha}} + D^{\phi}_c. \label{eq:collapse_Dphi_inflection}
\end{align}
As this expression has three unknown parameters compared to the 5 considered system sizes we first determine the best fit parameters $\tilde{D}(\bar{W}_0)$ and $D^{\phi}_c$ for various values of $\alpha$. An exemplary fit for $\alpha = 0.51$ is shown in Fig.~\ref{fig:SF-BG-scaling_U20}($b$). As the finite-size scaling at the critical point $W_c$ is independent of the scaling exponent $\nu$, we can further determine $W_c$ from the crossing points of the data sets for various system sizes if we only apply the scaling to the fractal dimension as shown in the inset of Fig.~\ref{fig:SF-BG-scaling_U20}($b$). This way we get the best candidates for the critical point $(W_c,D^{\phi}_c)$ as a function of $\alpha$ depicted in Fig.~\ref{fig:SF-BG-scaling_U20}($c$). To quantify the goodness of these fits we consider the adjusted coefficient of determination $\tilde{R}^2$ given in the inset of Fig.~\ref{fig:SF-BG-scaling_U20}($c$). For the considered range of $\alpha$ $\tilde{R}^2$ is almost constantly at its optimum but as the fractal dimension by definition is limited to $D_{\phi} \in [0,2]$ we find a fixed lower bound for $\alpha$ [see Fig.~\ref{fig:SF-BG-scaling_U20}($c$)].

For the final collapse we only have to consider $\alpha$ and $\nu$ in order to minimize the mean relative variance as a measure for the goodness of the collapse:
%\begin{widetext}
\begin{align} 
\chi_{D_{\phi}} = \sum_{L' > L} \sum_{\bar{W}} \frac{\left(\tilde{D}_{\phi,L} (\bar{W})  - \tilde{D}_{\phi,L'} (\bar{W})\right)^2}{2 \left[ \sigma\tilde{D}^2_{\phi,L} (\bar{W}) + \sigma\tilde{D}^2_{\phi,L'} (\bar{W})\right]} \frac{1}{\bar{C}_{D_\phi}}. \label{eq:alg_Dphi_mean_rel_dev} 
\end{align}
%\end{widetext}
Analogous to the finite-size scaling at the mobility edge, $\sigma\tilde{D}_{\phi,L} (\bar{W})$ are the standard errors of the mean determined from the disorder sampling while the normalization constant $\bar{C}_{D_\phi}$ is given by the total number of terms, $\bar{C}_{D_\phi} = \sum_{L' > L} \sum_{\bar{W}} 1$. In order to estimate the error of the obtained scaling exponents, this finite-size scaling procedure is repeated for 6 independent subsets of 10 disorder realizations each. For the best collapses we find $\chi_{D_{\phi}} = 0.53(9)$ corresponding to the mean-field scaling exponents $\alpha = 0.51(2)$ and $\nu = 1.77(11)$ while the critical point has $D_c^{\phi}/t = 1.956(7)$ at a disorder of $W_c/t = 3.1(4)$. We thus find a ground state transition point at $U/t = 20$ that is consistent with previous predictions of a superfluid to Bose-glass transition also at half-filling but in the limit $U/t \rightarrow \infty$ and for equal distributed box-disorder of the local potential $\epsilon_{\ell} \in [-W,W]$ \cite{Makivic1993,Zhang1995,Priyadarshee2006,AlvarezZuniga2015}.

\section{\label{sec:Experiment}Single particle correlations in a trap}

\begin{figure}[b]
 \centering
 %\captionsetup{width=0.95\textwidth}
 \includegraphics[width=0.99\textwidth]{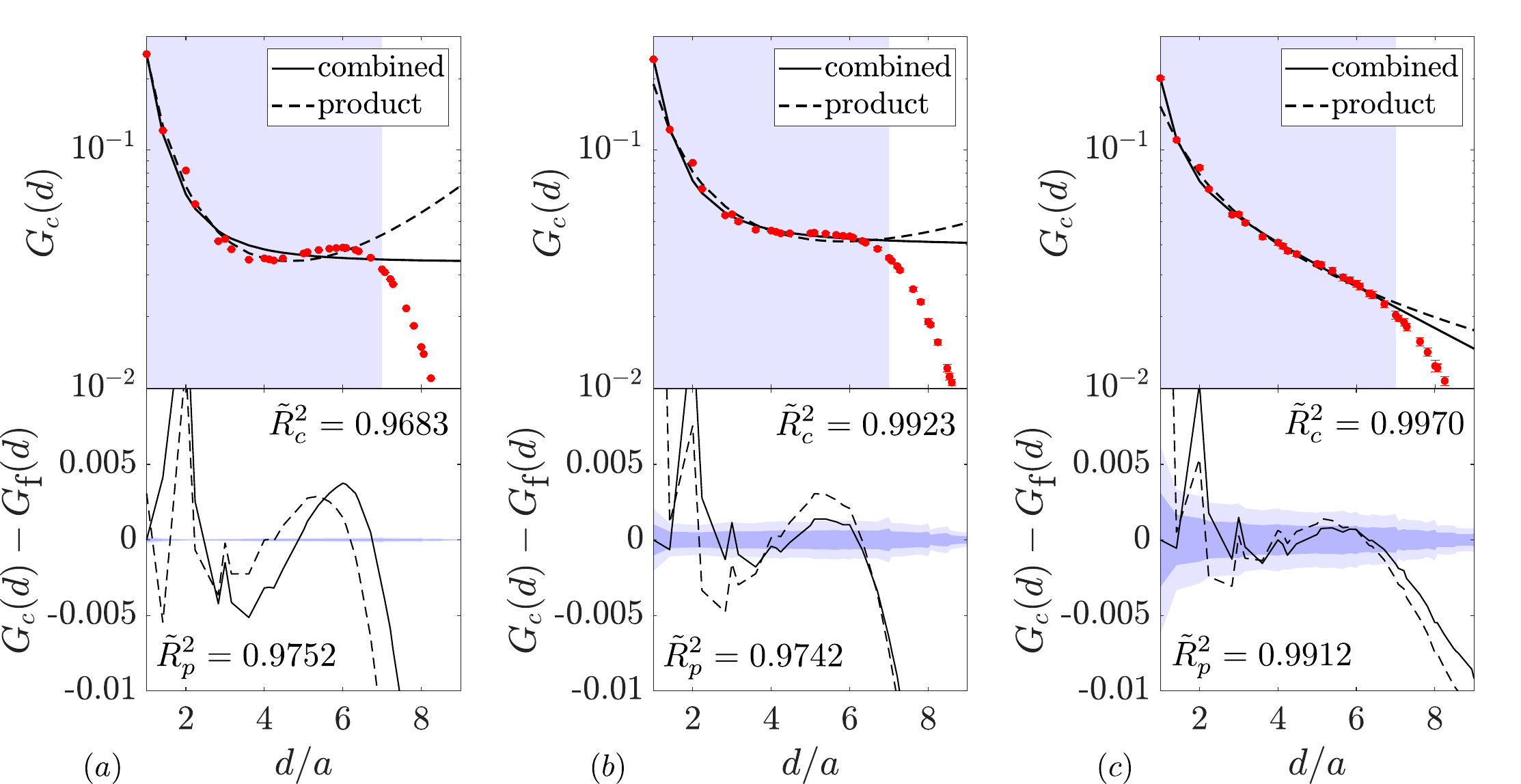}
 \caption{
Comparison of the product and linear combination fitting scenarios (see legend in first figure) for the radial connected correlation function $G_c(d)$ in a trapped system. Numerical data points are shown with an error corresponding to one standard deviation of the disorder sampling. Black lines are the best fits of each scenario (see legend), while the lower half of each panel depicts the residuals of the data $G_c(d)$ to each fit $G_\textrm{f}$ together with the adjusted coefficient of determination $\tilde{R}^2$. For $(a-c)$ from left to right the considered disorder values are $W/t \in \left\{ 0.4 , 2.8 , 6.5 \right\}$, respectively. The single shaded region in the background marks the fit region of $d/a \in \left[ 1,7 \right]$, while the two different shaded regions for the residuals correspond to one and two SD of the numerical data.
%\textit{Bottom:} Residuals of the best fits compared to the data (REPLACE LINES BY POINTS!, see legend). The shaded regions correspond to the ($1 \sigma$ and) $2 \sigma$ error of the numerical data.
 }
 \label{fig:trap_fit_scenarios}
\end{figure}

As in the experiment we consider a finite-size system in a harmonic trap \cite{Choi2016}, thus particles with a phase coherence over only a finite length scale, driven by the competition of repulsive Hubbard interactions and local disorder, may still span the whole system, forming a local BEC in the ground state. Only for vanishing disorder and interaction do all particles participate in the condensate mode, while a finite condensate order parameter is observed in the ground state even as both interaction and disorder are increased (with the typical wedding cake structure in the homogeneous case). For finite-size systems at the superfluid to Bose-glass transition the condensate order parameter also does not vanish at the critical point and beyond \cite{Meldgin2016}. In this confined system, it is thus reasonable to assume a crossover in the ground state where a fraction of the particles always exhibits long-range correlations with a short-distance algebraic decay, as in a condensate, and only the remaining fraction can be localized more strongly. So, as a first scenario for $G_c(d)$, where the distance $d$ is in units of the lattice spacing $a$, we assume the empirical linear combination form

\begin{align}
G_l(d) = A_l \left[ p \exp(\lambda - \lambda d) + \frac{1-p}{d^b} \right]. \label{eq:Greens_lin}
\end{align}
We fix the amplitude $A_l$ of this ansatz by the value $A_l = G_c(d=1)$ for nearest neighbors. Then $p$ corresponds to the relative contribution of the exponential decay with the decay constant $\lambda$ and the algebraic decay with the exponent $b$. We consider the latter three as fit parameters.

On the other hand, related work on Anderson localization on random regular graphs, where every site has $m+1$ nearest neighbors, suggests a product of exponential and algebraic decays instead \cite{Tikhonov2019},

\begin{align}
G_p(d) = A_p \frac{m^{1-d}}{d^{c}}. \label{eq:Greens_prod}
\end{align}
Due to the similarity of the effective quasiparticle Hamiltonian to such systems it might be considered a possible model theory for the non-interacting quasiparticles, with $m-1$ the effective number of nearest neighbors and $c$ the algebraic decay constant. Together with the amplitude $A_p$ we thus have three fit parameters, as well.

\begin{figure}[t]
 \centering
 %\captionsetup{width=0.95\textwidth}
 \includegraphics[width=0.48\columnwidth]{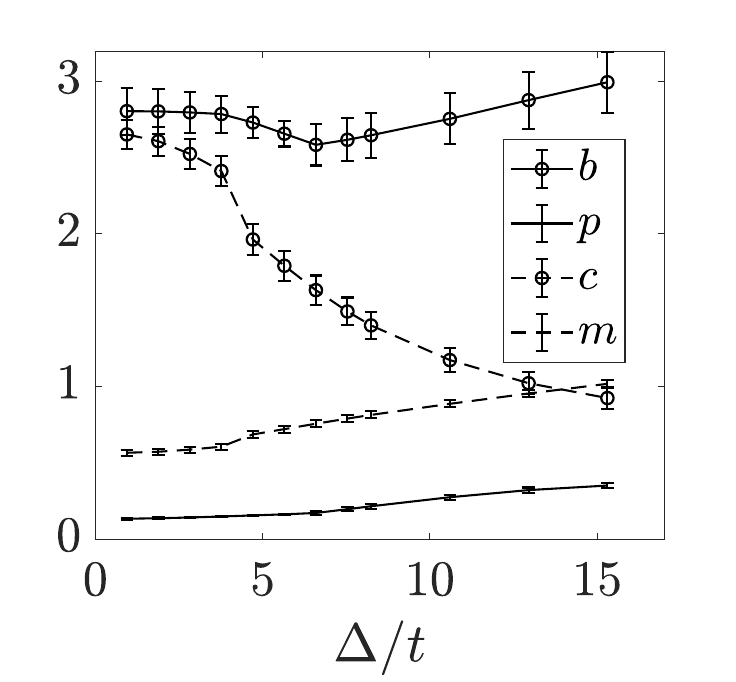}
 \caption{
Parameters (identified in the legend) of the two considered fitting scenarios of the connected Greens function as functions of the disorder strength. The parameters $(b,p)$ correspond to the linear combination~\eqref{eq:Greens_lin} while the parameters $(c,m)$ correspond to the product form~\eqref{eq:Greens_prod}.
 }
 \label{fig:trap_fit_parameters}
\end{figure}

In Fig.~\ref{fig:trap_fit_scenarios} we show exemplary best fits of these two scenarios to our numerical data for various values of the disorder $W/t \in [0.4 , 2.8, 6.5]$. In most cases the adjusted coefficient of determination $\tilde{R}^2$ for the product scenario is slightly worse compared to the linear combination form. Furthermore, only for strong disorder does the product form not diverge as $d \rightarrow \infty$. Except for the decay constant $\lambda$ of the exponential term in the linear combination form discussed in the main part, we show all the relevant parameters in Fig.\ref{fig:trap_fit_parameters}. Here we can see that the apparent unphysical behavior of the product form Eq.~\eqref{eq:Greens_prod} stems from $m < 1$. For the linear combination form Eq.~\eqref{eq:Greens_lin}, on the other hand, the exponent of the algebraic term remains constant within its errorbars, as one would expect for the assumed two components. Finally, the relative contribution of the localized component starts to increase precisely at the transition.

\section{Entanglement}\label{sec:entanglement}

In the following we show that the \ac{FOE}, while based on a mean-field representation, is able to capture effects of many-body entanglement in strongly correlated systems. To do so we first present a qualitative argument for the occurrence of non-trivial entanglement in this method and then discuss some numerical examples for the disordered Bose-Hubbard model central to this work.

A typical measure of entanglement is the von Neumann entanglement entropy $S_{\textrm{EE}}(\rho_A) = \textrm{Tr}\left[ -\rho_A \ln (\rho_A) \right]$ evaluated for the density operator $\rho_A$ of any subsystem $A$. For a closed system $S = A \cup B$ (with $A \cap B = \o$) one has $\rho_A = \textrm{Tr}_{B} \left[ \rho_S \right]$. In order for $S_{\textrm{EE}}$ to be non-zero, which signifies entanglement, the Schmidt rank counting the non-zero eigenvalues of $\rho_A$ has to be greater than one. We will now show this property for typical \ac{FOE} states. To perform the partial trace using the \ac{FOE} method we consider $\rho_{\textrm{QP}} \equiv | \psi_{\textrm{QP}} \rangle \langle \psi_{\textrm{QP}} |$ and an analogous form for any \ac{QP} added state $\beta_{\gamma}^{\dagger} | \psi_{\textrm{QP}} \rangle$. At first glance this ansatz seems to necessitate explicit knowledge of the state. In contrast, it is straightforward to consider the Gutzwiller bases to gain some insight into the structure of $\rho_{\textrm{QP}}$. Each of its matrix elements 
\begin{align}\label{eq:densmat_elements}
\left( \prod_{\ell} \langle i_{\ell} | \right) | \psi_{\textrm{QP}} \rangle \langle \psi_{\textrm{QP}} | \left( \prod_{\ell'} | j_{\ell'} \rangle \right) =  \langle \psi_{\textrm{QP}} | \left( \prod_{\ell'} | j_{\ell'} \rangle \right) \left( \prod_{\ell} \langle i_{\ell} | \right) | \psi_{\textrm{QP}} \rangle
\end{align}
is given by the correlations of local QP ground state excitations. The properties of $\beta_{\gamma}$ and $|\psi_{\textrm{QP}}\rangle$ guarantee that only correlations of even order in the Gutzwiller operators are non-vanishing. Thus, $\rho_{\textrm{QP}}$ is separable into two sectors, one consisting of even numbers and one consisting of odd numbers of local excitations as correlations coupling these sectors are of odd order in the Gutzwiller operators and thus vanish.

As $| \psi_{\textrm{QP}} \rangle$ represents a correction of the MF ground state, the relevant sector of $\rho_{\textrm{QP}}$ has to contain the MF state and is the even sector. 
Now we consider an arbitrary bipartition ($A$ and $B$) of a system $S$ consisting of at least two sites. Except for the trivial case $| \psi_{\textrm{QP}} \rangle = | \psi_{\textrm{MF}} \rangle$, coefficients then fall into two mutually distinct groups. Either, the corresponding basis vectors in the subspaces of $A$ and $B$ are both in the odd sector, or, they are both in the even sector. Thus there are always non-trivial submatrices of $\rho_{\textrm{QP}}$ where all involved eigenvector pairs of both subspaces $A$ and $B$ are mutually orthogonal. This implies that a Schmidt decomposition then yields a nontrivial Schmidt rank greater than one. Finally, as a single \ac{QP} added state $\beta_{\gamma}^{\dagger} | \psi_{\textrm{QP}} \rangle$ resides in the odd subspace of Gutzwiller excitations, one can apply the same argument with the only difference being that the paired subspaces of $A$ and $B$ are then always of opposite order. Therefore, we conclude that the \ac{FOE} predicts nontrivial entanglement properties of a strongly correlated system for the ground state as well as its many-body excitations \footnote{While, in principle, this argument can be repeated for any number of non-interacting FOE quasiparticle excitations, previously neglected interaction terms become increasingly relevant.}.

\begin{figure}[t]
 \centering
 %\captionsetup{width=0.95\textwidth}
 \includegraphics[width=0.99\columnwidth]{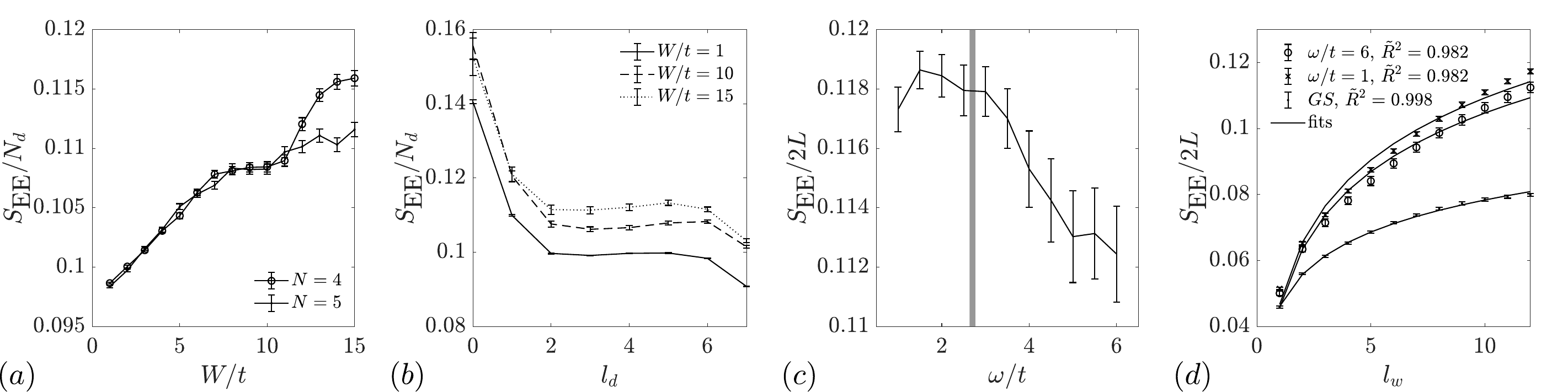}
 \caption{
Entanglement entropy in the disordered Bose-Hubbard model at half filling and for $U/t = 20$. For a system with $L=15$ $S_{\textrm{EE}}$, normalized by $N_d$ the number of sites on the surface of a region $A$ (see text) and for the \ac{FOE} ground state, is shown in $(a)$ as a function of disorder $W/t$ for the truncations $N=4$ and $N=5$ with $l_d=6$. $(b)$ depicts the dependence of $S_{\textrm{EE}}$ on $l_d$ the size of the region $A$ for fixed disorder values. The entanglement entropy $S_{\textrm{EE}}/2L$ for the \ac{QP} excitations of a 12-by-24 site system at $W/t = 6$ and for $N=4$ is shown in $(c)$ as a function of the \ac{QP} energy $\omega/t$ and $l_w=12$. The grey region denotes the \ac{ME} determined in the finite size scaling. In $(d)$ instead, the dependence of $S_{\textrm{EE}}$ on the width $l_w$ is compared for the ground state (GS) and fixed \ac{QP} energies (see legend). Black lines mark best fits to the logarithmic ansatz~\eqref{eq:log_fit}, with the quality of fit parameter given in the legend for each case. 
 }
 \label{fig:S_EE}
\end{figure}

Let us now quantify the amount of entanglement beyond this qualitative argument. The quadratic form of the approximate \ac{FOE} Hamiltonian $\hat{\mathcal{H}}^{(2)}_{\textrm{QP}}$ implies the applicability of Wick's theorem, as has been discussed for a closely related method \cite{Frerot2015,Frerot2016}. Thus, assuming the Gutzwiller operators to be exactly bosonic (see main text), the reduced density matrix $\rho_A$ can be reconstructed from the two-point correlations only. The relevant correlations are $C_{i\ell,j\ell'} \equiv \langle \sigma^{(i)\dagger}_{\ell} \sigma^{(j)}_{\ell'} \rangle$ and $D_{i\ell,j\ell'} \equiv \langle \sigma^{(i)}_{\ell} \sigma^{(j)}_{\ell'} \rangle$ for which one can show the relation
\begin{align}
\begin{pmatrix}
-\mathbb{1}-C^* & F \\
-F^* & C
\end{pmatrix}
=
U_A
\begin{pmatrix}
-\textrm{diag}(1+n_{\alpha}) & 0 \\
0 & \textrm{diag}(n_{\alpha})
\end{pmatrix}
U_A^{-1}
\end{align}
where $U_A$ is a Bogoliubov transformation and $n_{\alpha}$ is the occupation of the $\alpha$th mode of a quadratic Hamiltonian $\mathcal{H}_A$ fulfilling the identity $\rho_A = \exp(-\mathcal{H}_A)$. Thus the von Neumann entanglement entropy of $\rho_A$ can be obtained via
\begin{align}
S_{\textrm{EE}} = \sum_{\alpha} \left[ (1+n_{\alpha}) \ln (1+n_{\alpha}) - n_{\alpha} \ln (n_{\alpha}) \right].
\end{align}

This representation applies to any eigenstate of a quadratic Hamiltonian with anomalous hopping. Therefore, we can determine $S_{\textrm{EE}}$ for the ground state as well as the excitations. For the ground state we consider the case $U/t = 20$ and $W/t \in [1,15]$ for a system with $L=15$. We define the region $A$ as all sites $\mathbf{r}_{\ell}$ within a radius $l_d \geq |\mathbf{r}_{\ell}-\mathbf{r}_{c}|/a$ of a given site $\mathbf{r}_c$ where we have defined $|\mathbf{r}| \equiv |r_x| + |r_y|$. We average of 60 positions for $\mathbf{r}_c$ for each of the $N_r=10$ realizations we consider. Fig.~\ref{fig:S_EE}$(a)$ shows $S_{\textrm{EE}}$ as a function of the disorder strength $W/t$ for the truncations $N=4$ and $N=5$, as well as $l_d = 6$. One can see that $S_{\textrm{EE}}$ is independent of the truncation up to $W/t \approx 11$ indicating converged results for any disorder below this value. Furhtermore, the linear dependence of $S_{\textrm{EE}}$ on $W$ ends with a kink at about $W_c \approx 7t$. The position of this kink is close to the ground state phase transition discussed for the mean-field state in Sec.~\ref{sec:MF-BG}. In addition,  Fig.~\ref{fig:S_EE}$(b)$ shows that $S_{\textrm{EE}}/N_d$ as a function of $l_d$ is mostly constant for various values of the disorder strength. This is consistent with the expectation of an area law for the entanglement growth typical for the ground state. Deviations from a constant at large and small $l_d$ can be attributed to finite size effects within our periodic system.

Next we consider a system of $L \times 2L$ sites with periodic boundaries to analyze the entanglement in the \ac{QP} excitations $\beta^{\dagger}_{\gamma} | \psi_{\textrm{QP}} \rangle$ at $W/t = 6$ and $N=4$. To do so we define $A$ as strips of varying width $l_w \in [1,12]$ wrapped along the short direction. $S_{\textrm{EE}}(\omega)$ is then sampled for $N_r = 10$ realizations and over the 24 possible positions of the strip for each disorder realization, each time taking the excitation closest in energy $\omega_{\gamma}$ to $\omega$. As shown in Fig.~\ref{fig:S_EE}$(c)$ for a width of $l_w = 12$ sites, $S_{\textrm{EE}}(\omega)$ is approximately constant up to a critical excitation energy $\omega_c \approx 3t$ beyond which it decreases continuously. The position of this kink is consistent with the near-thermal \ac{ME} derived above (see Sec.~\ref{sec:ME_scaling}). Finally, by also considering $S_{\textrm{EE}}/2L$ as a function of the width $l_w$ we observe very different behavior in the ground state compared to the \ac{QP} excitations (see Fig.~\ref{fig:S_EE}$(d)$). On the one hand the ground state entanglement can be fit nearly exactly by the ansatz
\begin{align}\label{eq:log_fit}
\frac{S_{\textrm{EE}}(l_w)}{2L} = a \ln(l_w)+c, 
\end{align}
suggesting a logarithmic correction to the area law in the ground state. Such a correction is expected in a superfluid ground state \cite{Frerot2016} as the considered disorder strength is below the critical value. But, on the other hand panel $(d)$ shows that the best fits of the logarithmic ansatz for the excitations at $\omega/t 1$ (cross symbols) and $\omega/t = 6$ (circular symbols) have substantial systematic deviations and a much lower $\tilde{R}^2$ compared to the ground state. The (finite-size) entanglement growth of the excitations is thus inconsistent with a logarithmic growth suggesting a faster growth of $S_{\textrm{EE}}$ with the volume for fixed boundary area of the region $A$. The observed relation is approximately consistent with a square-root behavior, although an accurate prediction would require a greater range of system sizes beyond the scope of this discussion.

In conclusion, as discussed qualitatively and quantitatively in this section, the \ac{FOE} method is able to capture non-trivial entanglement both in the ground state and in its \ac{QP} excitations. In the ground state it predicts the commonly expected area law behavior, while it even predicts faster then area law (possibly even volume law) entanglement growth for the (near-)thermal \ac{QP} excitations in the vicinity of the \ac{ME}.

\end{document}